 \documentclass{aastex}

\newcommand{\rsun}{R$_\odot$}
\newcommand{\kms}{km\,s$^{-1}$}

\slugcomment{submitted to the Astrophysical Journal}

\shorttitle{CME Source Regions}
\shortauthors{Subramanian and Dere}

\begin{document}


\title{Source Regions of Coronal Mass Ejections}


\author{Prasad Subramanian\altaffilmark{1}}
\affil{Center for Earth Observing and Space Research, George Mason University,
Fairfax, VA 22030, USA}
\email{psubrama@iucaa.ernet.in}

\author{K.~P. Dere \today}
\affil{Naval Research Laboratory, Code 7660, Washington, DC 20375, USA}
\email{dere@halcyon.nrl.navy.mil}

\altaffiltext{1}{Present address: IUCAA, P.O Bag No: 4,
 Ganeshkhind, Pune 411007, India}


\begin{abstract}
Observations of the solar corona with the Large Angle Spectrometric 
Coronograph (LASCO) and Extreme ultraviolet Imaging Telescope (EIT)
 instruments on the Solar and Heliospheric Observatory
(SOHO) provide an unprecedented opportunity to study coronal mass
ejections (CMEs) from their initiation through their evolution out to
30 \rsun.  The objective of this study is to gain an understanding of
the source regions from which the CMEs emanate.  To this end, we have
developed a list of 32 CMEs whose source regions are located on the
solar disk and are well observed in EIT 195 {\AA} data during the period from solar
minimum in January 1996 through the rising part of the cycle in May 1998.  
We compare the EIT source
regions with photospheric magnetograms from the Michelson Doppler Imager (MDI)
 instrument on SOHO
and the NSO/Kitt Peak Observatory and also with H$\alpha$ data from
various sources. The overall results of our study show that 41\% of the
CME related transients observed are associated
with active regions and have no prominence eruptions, 44\% are associated with
eruptions of prominences embedded in active regions and 15\% are
associated with eruptions of prominences outside active regions. 
Those CMEs that do not 
involve prominence eruptions originate in active regions both with and without 
prominences. We describe 6
especially well observed events. These case studies suggest that active
region CMEs (without eruptive prominences) 
are associated with active regions with lifetimes between
11--80 days. They are also often associated with small scale emerging
or cancelling flux over timescales of 6--7 hours.  CMEs associated with
active region prominence eruptions, on the other hand, are typically
associated with old active regions with lifetimes $\sim$ 6-7 months.

\end{abstract}




\section{Introduction}
 
In this paper, we concentrate on CME source regions with a view to
gaining insight into the initiation mechanism of CMEs, especially in
relation to pre-existing coronal structures and the underlying
photospheric magnetic field. We confine our attention to CMEs observed
in LASCO white light images that have clearly identifiable on-disk
counterparts in EIT 195 {\AA} data that are well away from the solar
limb.  The photons detected by the EIT 195\AA\ channel are primarily
due to Fe XII formed at a temperature around 1.4 x 10$^6$K \citep{eitcal}.

We address such questions as:

\begin{enumerate}
\item
What are the source regions of CMEs?
\item
Are there any discernible characteristics of the source regions indicating
an inclination for producing CMEs?
\item
How are CMEs related to active regions and prominence eruptions?
\item
What is the short and long-term temporal evolution of CME producing
regions?
\end{enumerate}

Several recent papers have reported on instances of phenomena in the
lower corona in EIT 195 {\AA} data that can be linked directly with
CMEs observed with the LASCO coronograph.  Examples include the halo
CME associated with the EIT wave of May 12, 1997 \citep{thom98,plun98}
and the CMEs of January 25 1998 \citep{gopal98} and 23 December 1996
\citep{dere97}. \citet{del00} examine EIT 195 {\AA} data for a period
of 6 days between 1 and 6 November 1997 and find that of 17 ejections
observed in this data, 13 are related to white light CMEs in LASCO
data.

Several studies have been carried out on associating CMEs with other
kinds of solar activity such as flares and disappearing prominences.
\citet{munro79} used CME observations from Skylab data during 1973-1974
near solar minimum in conjunction with reports of associated activity
from the Solar Geophysical Data archives. They found that 78\% of the
CMEs observed were associated with other forms of solar activity.  Of
these, they found that 40\% of the CMEs observed are associated with
flares, while more than 70\% of CMEs are associated with eruptive
prominences (with and without flares).  \citet{webb87} examined SMM
data during 1980 near solar maximum and found that 66 \% of the CMEs
observed were associated with other forms of solar activity.  Of these,
they found that 68\% of the observed CMEs were associated with erupting
prominences, 37\% with H$\alpha$ flares and 47\% with X-ray events.
\citet{stcyr91} analyzed SMM data from 1984--1986, during the declining
phase of the solar cycle. They find that slightly less than half of the
CMEs have associations with other forms of solar activity. Of these,
they found that 76\% were associated with erupting prominences, 26\%
with H$\alpha$ flares and 74\% with X-ray events. \citet{gilbert00} have analyzed
H$\alpha$ data from the Mauna Loa Solar Observatory and LASCO data to study the
relationship between prominence activity and CMEs. Their data cover the period
February 1996 -- June 1998, which is very similar to the period covered in our study.
One of their principal findings is that 94\% of eruptive prominences discerned
in the H$\alpha$ data had CMEs associated with them.

We now review recent work examining the relation between CMEs and the
photospheric magnetic field.  \citet{feyn95} find a strong correlation
between newly emerging magnetic field and filament eruptions, which
they take to be proxies for CMEs.  \citet{luh98} employ potential field
extrapolations to determine the large scale coronal field from the
photospheric magnetic field. Their results suggest that although the
changes in the small scale photospheric magnetic field before and after
a CME might be indiscernible, CMEs could correspond to the opening up
of large scale coronal field lines.  \citet{lara00} examine 8 eruptive
events associated with active regions during 1997 and 1998. They find
that the total magnetic flux over an active region did not show
significant changes. However, they found significant changes in the
flux over small subregions of the overall active region associated with
the CMEs.  They also found that the changes in flux occur over
timescales of several hours to days.

Photospheric magnetic field configurations are the boundary conditions
for determining the coronal magnetic field which is disrupted during a
CME.  \citet{can99} have suggested that CME source regions often
correspond to structures resembling sigmoids in YOHKOH soft X-ray
images, that presumably trace highly stressed coronal magnetic field
configurations.

The issue of CME initiation is thus an issue of considerable
contemporary interest.  The combination of the highly sensitive white
light images provided by the LASCO coronograph, on-disk EIT images and
1.5 hour cadence (MDI) photospheric magnetograms provide a unique
opportunity to investigate the complex question of CME initiation.
This kind of comprehensive data set was not available when previous
studies, described above, of the associations between CMEs and disk
phenomena were made.  We address the question of CME initiation by
analyzing the source regions of CMEs observed on the solar disk in EIT
195 {\AA} images during the solar minimum period of January 1996--June
1998.

\section{Data Analysis Procedure}

Our main aim in this paper is to determine the source regions of CMEs
as seen in EIT 195 {\AA} images, understand the signatures of CMEs in
the low corona and compare them with underlying photospheric magnetic
field configurations. Line-of-sight magnetograms measure the
longitudinal component of the magnetic field. The magnetograms we use
from the NSO Kitt Peak and the MDI instrument are most sensitive near
disk center, since there is no foreshortening.  Accordingly, we have
searched the LASCO CME database
(\url{http://lasco-www.nrl.navy.mil/cmelist.html}) for the period
January 1996 - June 1998 to find CMEs in white light LASCO images that
have clear on-disk transients in EIT 195 {\AA} images associated with
them.  We search EIT 195 {\AA} images within approximately 20-25
minutes prior to the appearance of the CME in the LASCO C2 field of
view to see if there are identifiable transients that could correspond
to the initiation of the CME in the lower corona. Such signatures
include plasma/material motions, transient coronal holes/dimming
\citep{ster97,zarro99}, erupting prominences and EIT waves
\citep{thom99} and flares.  We have generally restricted ourselves to CMEs

\begin{itemize}
\item
that have unambiguous associations with on-disk phenomena observed in
EIT 195 {\AA} images, and
\item 
where there are high cadence (12-18m), full resolution (2.6
arc-sec/pixel) EIT 195 {\AA} images so that the initiation and source
region of the CME-related transients in 195 {\AA} EIT images is clearly
evident.
\end{itemize}

We have cataloged a total of 32 events that satisfy the above-mentioned
criteria during the solar minimum period of January 1996--June 1998
(Table 1).  Since this only a fraction of the CMEs observed by LASCO
during this time period, it is necessary to consider the effect that
our selection criteria might have on our final conclusions.  First, the
selected CMEs only refer to a specific time period near the rising
phase of solar cycle 22 and may not be tightly correlated with
other phases of the solar cycle.  However, the correlation of CMEs
observed with SMM with other forms of solar activity at solar maximum
\citep{webb87} and during the declining phase \citep{stcyr91} were 
similar and suggest that the correlations found in the
present study may be more generally applicable to other phases of the
solar cycle.  Our conclusions will certainly be affected by the
relatively small number of events in our selection but we expect the
basic conclusions to be broadly representative of this phase of the
solar cycle.  The criteria used to select the events we have examined
were arrived at for the purpose of a clear determination of the source
region and the short term photospheric magnetic field activity related
to the CME.  The signatures of CMEs in the EIT data are often subtle so
that the availability of high cadence EIT data is necessary for a clear
location of the source region. Since there were few high cadence 
EIT 195 {\AA} observations before
March 1997 because of telemetry restrictions, our data is biased
towards events from March 1997 onwards. The MDI photospheric magnetograms are
the only consistent set of data available for studying photospheric
field evolution at a reasonable cadence.  Since the magnetic
sensitivity of MDI drops off toward the limb, these data are best used
near disk center.  Consequently, we believe that the conclusions of
this study have been strengthened by limiting our list of CMEs to a
well-observed subset of all of the CMEs observed by LASCO during our
period of study.




\section{Results}

\subsection{Aggregate Results} 

We have recorded the heliographic coordinates of the source region of
the CME related transient in 195 {\AA} EIT images for each event.  We
have superimposed the heliographic positions of these source regions on
synoptic magnetograms available from NSO Kitt Peak.  If the outline of
the CME source region lies over an active region, we conclude that it
is active region related.  During this phase of the solar cycle, active
regions are usually well isolated and these identifications tend to be
straightforward. Two examples
of synoptic magnetograms with CME source regions superimposed are shown
in Figures 1a and 1b. We have carried out this procedure for all the 32
events in our catalog (Table 1), but we do not show all the relevant synoptic
magnetograms for lack of space.

We have employed a variety of sources to ascertain whether a CME is associated with
an erupting prominence. H$\alpha$ data from the Big Bear Solar Observatory (BBSO)
and the Observatory of Paris at Meudon and some other sources
like the Holloman Air Force Base and the Hiraiso Solar Terrestrial Research
Center often help to show whether a prominence
existed near the source region of the CME related transient.
Prominence eruptions are occasionally clearly
evident in EIT 195 {\AA} data as cool prominence material seen in absorption
or as bright ejecta seen in emission. We have also examined
EIT 304 {\AA} data for prominences, but this is usually not very useful because
the time interval between images is typically as large as 6 hours. Erupting
prominences are also occasionally clearly evident in LASCO C2 data as highly
structured material. We have also used Solar Geophysical Data reports of prominence
eruptions.

  This exercise allows us to place the
source regions of CMEs into the following broad categories (see Table 1):
\begin{enumerate}
\item
CME source regions that are co-spatial with active regions (AR
category). Prominences are sometimes embedded in these active regions, but
there is no evidence to suggest that any part of it erupted with the CME.
\item
CME source regions that are related to eruptions of prominences that
were embedded in active regions (AR+EP category). In some cases, the entire 
embedded prominence
erupts as part of the CME, while in others only part of the prominence
does so. The prominences in
this category are often embedded in old, decaying active regions.
Such regions might not necessarily be assigned a NOAA active region
number. We have labelled some events in this category as AR+(EP). These are events
for which evidence for eruption of a prominence embedded in the active region is
not very conclusive.
\item
CME source regions that are related to eruptions of
prominences outside active regions (EP category). 
The prominences in this category occur in
quiet regions where the magnetic fields are relatively weak.
\end{enumerate}

The association of CME source regions with active regions  (events in
category 1) is made by visual inspection of EIT 195 {\AA} images and
synoptic magnetograms.  The association with erupting prominences, on
the other hand, is made by inspecting data from various sources as
described above.  Table 1 gives a short summary of all the events
considered in this study. As summarized in Figure 2, we
find that 13 (41\%) of the cataloged events fall in the AR category, while
14 (44\%) are in the AR+EP category and 5 (15\%) are in the EP category. We have
included events labelled as AR+(EP) in Table 1 in the count for AR+EP events. 

\citet{feyn95} find that quiescent filament eruptions (which they take
to be proxies for CMEs) are strongly correlated with emergence of new
magnetic flux over several days preceding the eruption. Our findings in
this respect are different.  As shown in Figure 2, 14 of the 32 events
on our list are in the AR+EP category, which was not included in the
study by \citet{feyn95}, while 5 of the 32 events on our list are in
the EP category. The CMEs in the EP category arise from eruptions of
prominences outside active regions.  
There is no evidence for significant
large-scale emerging
magnetic flux for any of the 5 events in the EP category.  
Nevertheless, we have observed quiescent filament eruptions
that were associated with newly emerged magnetic flux, for example, the
west limb CME on 1996 July 10.  This event was not included in our
study because of the lack of high cadence EIT data.

Of the 14 events in the AR+EP category, we find that 4 are associated
with emerging magnetic flux.  For 3 of these 4 AR+EP events, the flux
emerges over timescales of $\sim$ 7--27 days. Only 1 of these 4 events
shows flux emergence over a timescale of $\sim$ 1 day. Our conclusions regarding
emergence or cancellation of magnetic flux (or the lack thereof) are based on detailed
visual inspection of MDI magnetograms before and after the event. 

\subsection{Case Studies}

Sorting the various CMEs into 3 categories provides a useful overall
description of the source regions of CMEs.  Nevertheless, such a broad
characterization covers over much of the complexity of the nature of
the CME source regions and the CME initiation.  We have therefore
selected 6 especially well observed events from the list of 32 events
for closer examination (Table 2). The
principal criterion used in the selection of this subset of 6 events
was the availability of high cadence EIT 195 {\AA} images.  For these
events, we have examined the EIT 195 {\AA} images in detail and made
detailed visual comparisons of the EIT 195 {\AA} images with full disk
photospheric magnetograms from the MDI instrument in order to determine
the spatial and temporal relationship between the photospheric magnetic
field evolution and the CME-related transient in EIT 195 {\AA} images.
We have also examined X-ray light curves from GOES data to determine if
there are flares associated with each of these events. In particular, we have referred
to the X-ray flare classification based on GOES X-ray data in the 0.1-0.8 nm band that
is used in the Solar and Geophysical Data reports.

\subsubsection{The CME of 01 April 1997}

This is an example of an active region (AR) associated CME and is
associated with active region 8026 (lower panel of Figure 3a) during
Carrington rotation 1921. The active region was absent on the previous
rotation, and emerged on the back side of the sun after 14 March 1997.
Since it is an isolated active region, loops associated with this AR
can be discerned from LASCO C1 images as early as 22 March 1997, well
before the AR reaches the East limb. The AR rotates into view in NSO
Kitt Peak magnetograms close to the East limb during rotation 1921 on
27 March 1997 (S23E78).  The CME of 01 April 1997 occurs when the AR is
at S24E20 (Figure 3a). Following the CME, the active region weakens and
disperses, but retains its identity well into rotation 1922. It was
last seen near the west limb on 13 May 1997 during rotation 1922 in MDI
magnetogram data and does not reappear at the east limb on 26 May 1997.
In summary, the active region, with which the CME of April 1 1997 is
associated, appeared between 14 March 1997 and 22 March 1997 and
disappeared between 13 May 1997 and 26 May 1997. The lifetime of the
active region is between 53 and 75 days.  The CME occurred early in the
life of the active region.
 
AR 8026 consists of two compact regions of opposite magnetic
polarities, with a parasitic negative polarity embedded in the positive
polarity region (Figure 3a). On 01 April 1997, the sunspot
corresponding to AR 8026 is assigned a $\beta$ classification.  The
parasitic negative polarity region is evident in the MDI magnetogram
images from as early as 23:59 UT on 1997/03/31.  Upon comparing the two
upper panels of Figure 3a, it is evident that that the parasitic
negative polarity evolves from being embedded in the positive polarity
region at 08:03 UT to forming a lane between two segments of the
positive polarity region by 14:27 UT. This trend continues, and we find
that the negative parasitic polarity lane has elongated further by
17:39 UT.  On 1 April 1997, the MDI magnetograms show a fair amount of
small-scale flux emergence and cancellation, particularly along the
neutral line and the western edge of the active region.

Near the time of the CME, the first sign of activity in EIT 195 {\AA}
images is a brightening at 13:25 UT apparently at the top of the arch
connecting the dominant polarities of AR 8026 (upper panel of Figure
3b).  This is followed by a brightening at 13:46 UT near the eastern
leg of the bipolar arch, around the location of the parasitic polarity
(lower panel of Figure 3b).  The brightening near the parasitic
polarity is clearly evident in the straight EIT 195 {\AA} image at
14:00 UT (Figure 3c). This sequence of events is accompanied by an M1.9
X-ray flare that starts at 13:25 UT, peaks at 13:48 UT and ends at
14:01 UT.  An EIT wave is observed to emanate from the
vicinity of the active region at 14:00 UT (lower panel of Figure 3b).
The northwestern front of this disturbance is the only one that can be
clearly identified as an outwardly propagating feature in successive
images.  It starts out with a velocity of $\sim$ 100 \kms\ and
propagates with an acceleration of $\sim$ 70 ${\rm m}{\rm s}^{-2}$.  By
the time it reaches the northwest edge of the disk, it has attained a
speed of $\sim$ 300 \kms.  A stationary dimming region is observed to
the southwest of the active region.  It is manifested as a deepened
darkening at approximately the same location in the running difference
EIT 195 {\AA} image at 14:18 UT.  We also observe a stationary
transient dimming at 14:18 UT to the southeast of the active region.
We have checked constant base difference EIT 195 {\AA} images to verify
that these features are not artifacts of the running difference
technique.  In summary, the activity observed in EIT 195 {\AA} images
consists of a flare(s), an EIT wave propagating towards the northwest
and dimming regions to the southeast and southwest of the active
region.

The LASCO C2 observations of the CME are shown in Figure 3d.
These running difference LASCO C2 images show a wide, bright mound on
the east limb appearing simultaneously with a small bright ejection on
the west limb at 15:18 UT. The wide front on the east limb travels
outward at a speed of 296 \kms\
(\url{http://lasco-www.nrl.navy.mil/cmelist.html}).  The appearance of
this CME is considerably different from those that are typically
ejected from the limb but is consistent with a CME ejected from the
solar disk in the general direction of SOHO. The simultaneous ejection
from the two limbs suggests that this CME could be a toroidal CME
\citep{bru98} propagating towards the earth.  It is worth noting that
the EIT wave propagates to the northwest, whereas the CME observed in
LASCO C2 images is concentrated along the equator, with the brightest
feature on the east limb.  Consequently, the propagating front in EIT
195 {\AA} data is probably not directly related to the white light CME.
However, the visibility of the EIT wave on the disk could be
non-uniform, and the northwestern feature could be the most evident
one. Furthermore, CMEs originating near the limb that start out
propagating at relatively large angles to the equator eventually tend
to follow the large-scale dipole field and bend towards the equator as
they propagate out.
 
\subsubsection{The CME of 21 October 1997}

This event is classified in the AR+EP category.  The source region of
this CME is the old, decaying active region complex that includes AR
8097 and the magnetic fields to its northeast (Figure 4a).  Synoptic
magnetograms from the NSO Kitt Peak reveal that this active region
complex emerges during Carrington rotation 1926. It then undergoes
considerable evolution, but does not disappear until Carrington
rotation 1932.  The lifetime of the overall complex is thus $\sim$ 6
months, and this CME occurs $\sim$ 3 months after the complex first
appeared. The emergence of AR 8097 represents the emergence of new flux
into the original active region complex.  AR 8097 was not present at
the west limb during rotation 1927 on 02 October 1997. It grew on the
backside of the Sun and rotated into view at the East limb in Kitt Peak
NSO magnetogram data on 16 October 1997 (Carrington rotation 1928).
There was evidence for loops associated with this active region in EIT
195 {\AA} images when this active region was behind the east limb, on
14 October 1997. We therefore conclude that AR 8097 emerged between 02
October 1997 and 14 October 1997.  Synoptic magnetograms from the NSO
Kitt Peak reveal that this active region decayed slowly and disappeared
between 13 January 1998 and 12 February 1998.  There is evidence for
small-scale flux emergence, cancellation and decay in the
photospheric magnetic field (Figure 4a), but the relation of these
changes to the activity in EIT 195 {\AA} images is not clear. In
summary, the CME of 21 October 1997 occured during the rising phase of
new flux injected into an old, decaying active region complex.

The activity observed in the EIT 195 {\AA} images includes a bright EIT
wave accompanied by propagating density depletions, a possible
prominence ejection, ejection of bright material to the southeast and
flare-associated brightenings. The EIT 195 {\AA} images available
around the time of this event suggest that an EIT wave is initiated
from AR 8097 between 16:18 UT and 17:34 UT.  The top panel of Figure 4b
shows the pre-initiation state of the EIT wave at 16:18 UT, while the
bottom panel shows the EIT wave in progress at 17:45 UT. Unfortunately,
there is no EIT 195 {\AA} data between 16:18 and 17:34.  The wave is
observed to spread in a semi-isotropic manner, propagating
preferentially towards the south. The most pronounced wave-associated
darkenings (which could be interpreted as density depletions) are
observed towards the north and south of the active region. The
propagating depletion wave is also accompanied by material ejection
that is manifested as the bright N--S feature at the southwestern
boundary of the dark region in the lower panel of Figure 4b. This event
is accompanied by a C3.3 flare that starts at 17:00 UT, peaks at 17:54
UT and ends at 18:16 UT.  At 17:34 UT we observe a bright linear
feature near the magnetic neutral line that we interpret as the
prominence activation.  Later, at 17:45 UT, we observe a double-ribbon
flare-associated brightening on either side of the prominence and at
the footpoints of the post-flare loops which are later seen in EIT 195
{\AA} images at 18:12 UT (Figure 4c).

There is some evidence for an erupting prominence from EIT 195 {\AA}
images; a dark feature embedded in the active region is seen to travel
outward after 16:18 UT.  Also, at 17:34 a bright linear feature is seen
along the neutral line between the two flare ribbons that are seen
later in the flare.  Furthermore, an H$\alpha$ 6563 {\AA} image from
the Big Bear Solar Observatory taken at 15:26 UT on 21 October 1997
reveals an embedded filament in the active region. The corresponding
image taken at 15:45 UT on 22 October 1997 shows that the embedded
filament has largely disappeared. It is therefore quite likely that
this CME is associated  with the eruption of a part of this embedded
prominence.  Consequently, we have place this CME in the active region
with prominence (AR+P) category.

This sequence of events is associated with a halo CME at 18:18 UT in
LASCO C2 data (Figure 4d). The CME front is brightest on the east limb,
although a faint front extends all over the northen hemisphere.  The
southwestern front of the halo is somewhat delayed with respect to the
other fronts. The leading edge on the east limb travels outward at a
speed of 465 \kms, with no appreciable acceleration.

\subsubsection{The CME of 20 January 1998}

The source of this CME is placed in the AR category.  The source region
of this CME is the compact active region 8135 (lower panel of Figure
5a). On 20 January 1998, AR 8135 is associated with a $\beta$ sunspot.
There was no evidence for this active region at the west limb on 01
January 1998 during rotation 1930. AR 8135 appears near the East limb
in Kitt Peak magnetogram data on 13 January 1998. EIT 195 {\AA} images
show evidence for this active region until as far as 24 January 1998.
Since this is a fairly weak active region, it is difficult to discern
if it survives beyond 24 January 1998, either from EIT 195 {\AA} or
from MDI magnetogram data. It is, however, clear that this active
region does not appear at the East limb during rotation 1931 on
February 6 1998.  We therefore conclude that AR 8135 appeared between
01 and 13 January 1998 and disappeared between 24 January 1998 and 6
February 1998. The lifetime of AR 8135 is fairly short, between 11 and 36 days.
The CME of 20 January 1998 occurred roughly midway during the lifetime
of the active region.

A comparison of the top left and top right panels of Figure 5a reveals
flux cancellation to the south of the dominant polarities in the active
region over a 6 hour timescale.  There was no evidence for an erupting
prominence in this active region.  This event is therefore an example
of those in the active region (AR) category.

The EIT 195 {\AA} images of Figure 5b show the development of a
transient dumbell shaped depletion region behind faint loop-like
structures.  The depletion region starts forming at 20:53 UT and
deepens by 21:05 UT and extends to the north and south of AR 8135.
Constant base difference images also show this depletion region.  The
faint loop-llike structures appear to connect opposite magnetic
polarity  with high field strength as seen in the MDI images.  Unlike
some of the events previously described, the outward propagation of the
depletion region is very limited.  This suggests that the dimming is
not an EIT wave.

The activity seen in EIT 195 {\AA} images is associated with the a
faint CME on the west limb.  The first signs of this CME in LASCO C2
images are observed at 22:03 UT. Figure 5c shows the successive stages
of the CME as it develops and propagates outward.  The leading edge of
the CME travels outward at a speed of 253 \kms\ with no appreciable
acceleration (\url{http://lasco-www.nrl.navy.mil/cmelist.html}).  As
with the AR event of 01 April 1997, the direction of motions observed
in EIT 195 {\AA} images do not appear to be well correlated with the
CME seen in LASCO C2 images; while the density depletion formed in the
southwest, the CME was observed over the west limb. However, as
mentioned before, it is possible that the large scale dipole field
channeled the CME that originated in the lower corona towards the
equator.  There is no evidence of a flare associated with this event. A
C1.5 flare occurred between 19:32 UT and 19:46 UT, but it was evidently
not associated with the sequence of events related to this CME.

\subsubsection{The CME of 25 January 1998}

This is a fairly complex event that has been classified in the AR+EP
category.  This CME has also been studied in some detail by
\citep{gopal98}.  It involves a newly emerged active region AR 8145 and
decaying active region flux that contains a quiescent filament.  The
decaying active region complex shown in Figure 6a has a lifetime of
$\sim$ 7 months, first appearing during Carrington rotation 1927 and
disappearing by rotation 1933.

During the previous rotation (1931), there was no evidence for 
AR 8145 near the west limb on 11 January 1998. AR 8145 grew on
the backside of the sun and came into view during rotation 1932 near
the east limb in MDI magnetogram data on 24 January 1998. MDI
magnetogram data shows that AR 8145 retains its identity until it is
close to the west limb on 03 February 1998 during rotation 1932, but
does not appear on the east limb on 16 February 1998 during the
subsequent rotation. AR 8145 appeared between 11 January and 24 January
1998 and disappeared between 03 February and 16 February 1998.
Its lifetime is thus between 11 and 37 days. Either way, it is evident
that the CME of 25 January 1998 occurred during the rising phase of the
active region.  Since AR 8145 is fairly close to the limb around the
time of the CME, projection effects limit our ability to discern
short-term changes in the photospheric magnetic field.

This CME occurs $\sim$ 6 months
after the overall complex first appeared, during rotation 1932.  Unlike
the AR+EP CME of 21 October 1997, this CME does not seem to be
associated with emerging flux into the old, decaying active region
complex.  However, the EIT 195 {\AA} images of Figure 6c, which show
the latter stages of development of the CME suggest that the
large-scale fields associated with this CME are rooted in AR 8145
towards the east of the decaying active region complex.

The EIT 195 {\AA} images of Figure 6b show an EIT wave and an erupting
prominence starting at 14:19 UT. Figure 6c shows the latter stages of
evolution of the erupting filament and the associated EIT wave. The EIT
wave is comprised of a depletion region behind a bright front. As
before, we have examined constant base difference images to verify that
the depletion and brightenings are not merely artifacts of the running
difference technique.  At around 14:32 UT, the depletion region
develops and propagates to the southeast. By 15:02 UT, the bright front
of the EIT wave have covered a substantial portion of the solar disk.
This sequence of events is accompanied by a C1.1 X-ray flare that
starts at 14:29 UT, peaks 15:12 UT and ends at 17:00 UT.  The
post-flare loop brightenings evident in Figure 6c are primarily
concentrated across the neutral line of the old active region fields.

The EIT He II 304\AA\ image at 13:47, one of 4 synoptic 304\AA\ images
obtained on that day, shows a filament that gives the impression of
lifting off from the old active region neutral line.  The time series
of EIT 195\AA\ images indicates that the hot ejecta that also
participate in the CME observed in LASCO C2 originate in the newly
emerged AR 8145 at 14:19.  Other ejecta also appear from the south-west
of the old active region flux.  It is possible that Feynman and Martin
would have classified this event as a quiescent filament eruption that
was associated with the emergence of new, nearby magnetic flux.  The
orientation of the polarities of the new flux is likely to result in a
reconnection of magnetic flux between the new and the old flux system.

A very bright CME is observed above the east edge of the LASCO C2
occulter at 15:26 (Figure 6d).  This CME is a partial halo but the halo
signatures are most prominent to the south.

\subsubsection{ The CME of 03 May 1998}
 
The source region of the 1998 May 3 CME is of the AR+EP type.  The source
region AR 8210 is the prolific source of several CMEs between 27 April
1998 and 06 May 1998 during rotation 1935.  It was absent at the west
limb on 16 March 1998, during rotation 1933. It grew on the backside of
the sun and rotated into view at the east limb on 29 March 1998 during
rotation 1934.  AR 8210 rotates into view at the east limb on 25 April
1998 during rotation 1935. AR 8210 is close to the west limb on 08 May
1998 during rotation 1935.  A considerably weakened version of this
active region rotates into view on the east limb during rotation 1936
on 21 May 1998. It disappears during its passage across the disk
between June 2 and June 3 1998.  In summary, the active region appeared
between 16 March and 29 March 1998 and disappeared around 3 June 1998.
The overall active region lifetime is thus $\sim$ 65-79 days.  This is
a prolific active region that produced several CMEs and large 
flares during its transit 
across the disk during rotation
1935 (Table 3). All the CMEs that originate
from this active region between 27 April 1998 and 06 May 1998 (Table 3)
take place roughly during the mid-phase of this active region. Figure
7a shows MDI magnetogram images of AR 8210 around the time of the CME.
The lower panel of Figure 7a shows the position of AR 8210 on the solar
disk, while the upper panels show closeups of the active region. On 03
May 1998, AR 8210 is associated with a $\beta$--$\gamma$--$\delta$
sunspot.  It can be seen that there is small scale flux cancellation in
the areas marked by the black arrows over a timescale of $\sim$ 6 hours
prior to the initiation of the CME of 03 May 1998.  The general level
of small-scale flux emergence and cancellation in this active region is
considerably enhanced over other source regions discussed in this
paper.  The strength of this activity may be related to high rate of
CME production.

On May 03 1998, the first sign of activity in EIT 195 {\AA} data is a
brightening in the central part of AR 8210 at 21:25 UT. Bright prominence material
is also ejected to the north in EIT 195 {\AA} data at 21:25 UT. This is
followed by propagation of a density depletion towards the north
accompanied by material ejection at 21:39 UT (upper panel of Figure
7b).  The brightening of AR 8210 continues, and the patchy transient
density depletion covers most of western half of disk by 21:57 UT
(lower panel of Figure 7b).  This sequence of events is accompanied by
an M1.4 flare that starts at 21:12 UT, peaks at 21:29 UT and ends at
21:49 UT. The CME is first observed in the LASCO field of view over the
northwest limb at 22:02 UT (Figure 7c), propagating outwards with an
initial speed of 639 \kms\
(\url{http://lasco-www.nrl.navy.mil/cmelist.html}).  In the C2 field of
view, the CME propagates to the northwest even though its active
region source lay in the southern hemishpere.

\subsubsection{The CME of 19 May 1998}

This is an example of a CME that falls in the EP category (Figure 2).
The prominence is formed along the neutral line of the relatively weak,
extended magnetic fields to the west of AR 8222 (Figure 8a). These
fields appear to be the dispersed remnants of AR 8203 which emerged
during the previous rotation (rotation 1935).  We studied the active
region and the weaker magnetic fields to its west for a period of 5
days before the CME.  We found no evidence for significant large-scale
emerging or cancelling flux near the location of the prominence over
this timescale.  A comparison of the upper and lower panels of Figure
8a shows that there are no significant small-scale changes in magnetic
flux over timescales of a few hours either.

The EIT 195 {\AA} images of Figure 8b and 8c show the prominence
eruption starting at 09:34 UT. The entire prominence disappears in
unison.  The eruption is well in progress by 09:56 UT.  In particular,
the erupting prominence is clearly evident in the straight EIT 195
{\AA} image of Figure 8b. The prominence eruption results in the CME in
LASCO C2 data shown in Figure 8d. The prominence is clearly evident as
the bright filamentary structure in the right panel of Figure 8d.
There was a C4 flare associated with this CME.

\section{Discussion and Conclusions}

We have surveyed 32 CMEs during January 1996 - June 1998
whose source regions are well observed on the solar
disk in EIT 195 {\AA} images. All these CMEs 
occurred near solar minimum during the rising phase of cycle 21, 
and are tabulated in Table 1.
As depicted in Figure 2, we find that 13 (41\%) of these events are 
associated with active regions without
prominence eruptions, 14 (44\%) are associated with 
eruptions of active region prominences, and 5 (15\%) are associated with
eruptions of prominences outside active regions. 
Of the 14 CMEs that are associated with
eruptions of active region prominences, 
4 show evidence for emerging magnetic flux over 
several days before the CME.

These statistics contrast
with those obtained from Skylab and SMM observations
\citep{munro79,webb87,stcyr91} which find that a significantly larger fraction
of CMEs are associated with prominence eruptions. However, the differences
between these studies should be kept in mind: the current study considers
only those CMEs that have on-disk counterparts in EIT 195 {\AA} data.
On-disk CMEs are considered because we wish to
compare the CME source regions with the underlying
photospheric magnetic field, and longitudinal magnetograms are more
reliable when observed close to disk center. The studies made with
Skylab and SMM data, \citep{munro79,webb87,stcyr91} on the other hand, 
mostly consider CMEs seen on the limb. None of these studies
had accompanying full-disk EUV data. The association of CMEs with
prominences in these studies was made by considering prominences within 
specified latitudinal and
longitudinal ranges of the CME locations as inferred from white light
coronograph data.  

We have described 6 especially well observed 
events from our overall catalog
of 32 CMEs that have identifiable on-disk signatures in EIT 195 {\AA} images.
The principal criterion used in selecting these 6 events was the availability
of high resolution, high cadence EIT 195 {\AA} data. We have discussed
these events in some detail in \S 3.2.
We have abstracted some salient characteristics from these
descriptions in Table 2. It suggests that
active region CMEs (AR) are associated with 
active regions with lifetimes ranging from $\sim$ 11-80 days. These
CMEs occur either during the rising phase or mid-life of the active regions.
AR CMEs are also often associated with growing parasitic polarity
magnetic fields or with cancelling magnetic flux over timescales of
$\sim$ 6 hours near the initiation time of the CME. It should be emphasized
that such changes in magnetic flux take place over timescales that are much
longer than the timescale over which the CMEs are initiated. While these
changes in the photospheric magnetic field 
take place over timescales of around 6-7 hours, the CME
initiation processes typically take place over timescales of $\sim$ 30 minutes
- 1 hour. Furthermore, there is no particularly distinguishing feature
that sets these changes in the photospheric magnetic field apart from
similar changes in other non-CME producing active regions, based on a fairly
non-critical study.
CMEs that result from
erupting active region prominences (AR+EP), 
are associated with old, decaying active regions 
with lifetimes greater than 6 months.

\acknowledgments

PS acknowledges several useful discussions with Dr. Angelos Vourlidas.
We acknowledge several useful comments and suggestions from the anonymous referee.

This work was made possible with funding from NASA.  SOHO is a project
of international cooperation between ESA and NASA. The SOHO/LASCO data
used in this paper are produced by a consortium of the Naval Research
Laboratory, Max-Planck-Institut fuer Aeronomie (Germany), Laboratoire
d'Astronomie (France), and the University of Birmingham (England). We
gratefully acknowledge the SOHO SOI/MDI team for use of data from the
MDI instrument.  We thank the Big Bear Solar Observatory and the Observatory
of Paris at Meudon for making their data available on the SOHO synoptic data
webpage. NSO/Kitt Peak data used here are produced
cooperatively by NSF/NOAO, NASA/GSFC, and NOAA/SEL.




\clearpage



\figurenum{1a}
\figcaption[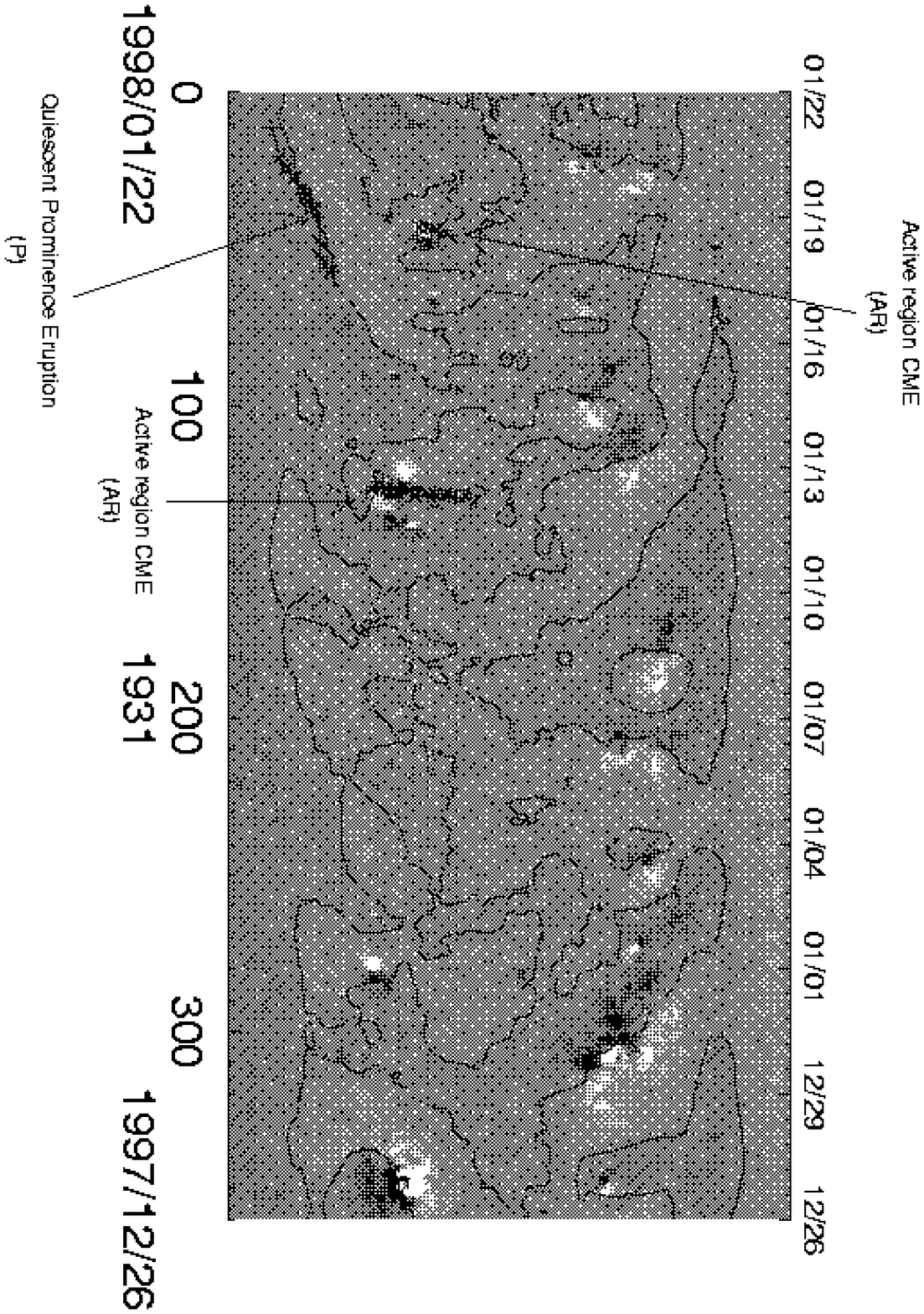]{The background is the synoptic magnetogram from the Kitt
Peak NSO for
Carrington rotation 1931. The on-disk CMEs observed in EIT 195 {\AA}
data during this rotation are superimposed on the magnetogram. \label{f1a}}

\figurenum{1b}
\figcaption[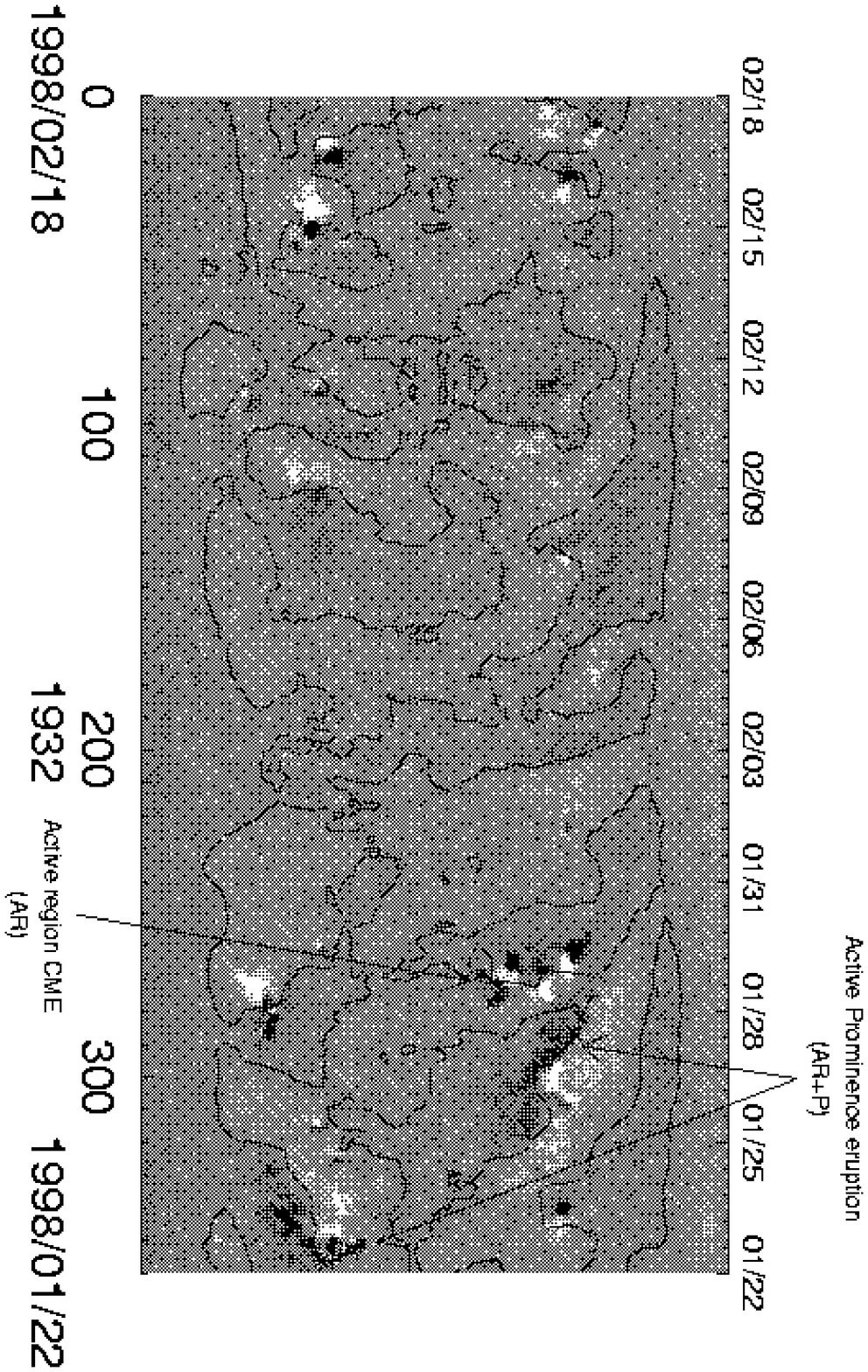]{The background is the synoptic magnetogram from the Kitt
Peak NSO for
Carrington rotation 1932. The on-disk CMEs observed in EIT 195 {\AA}
data during this rotation 
are superimposed on the magnetogram. We have examined such magnetograms
for rotations 1915-1935. We show only two examples here in order to
conserve space.
\label{f1b}}

\figurenum{2}
\figcaption[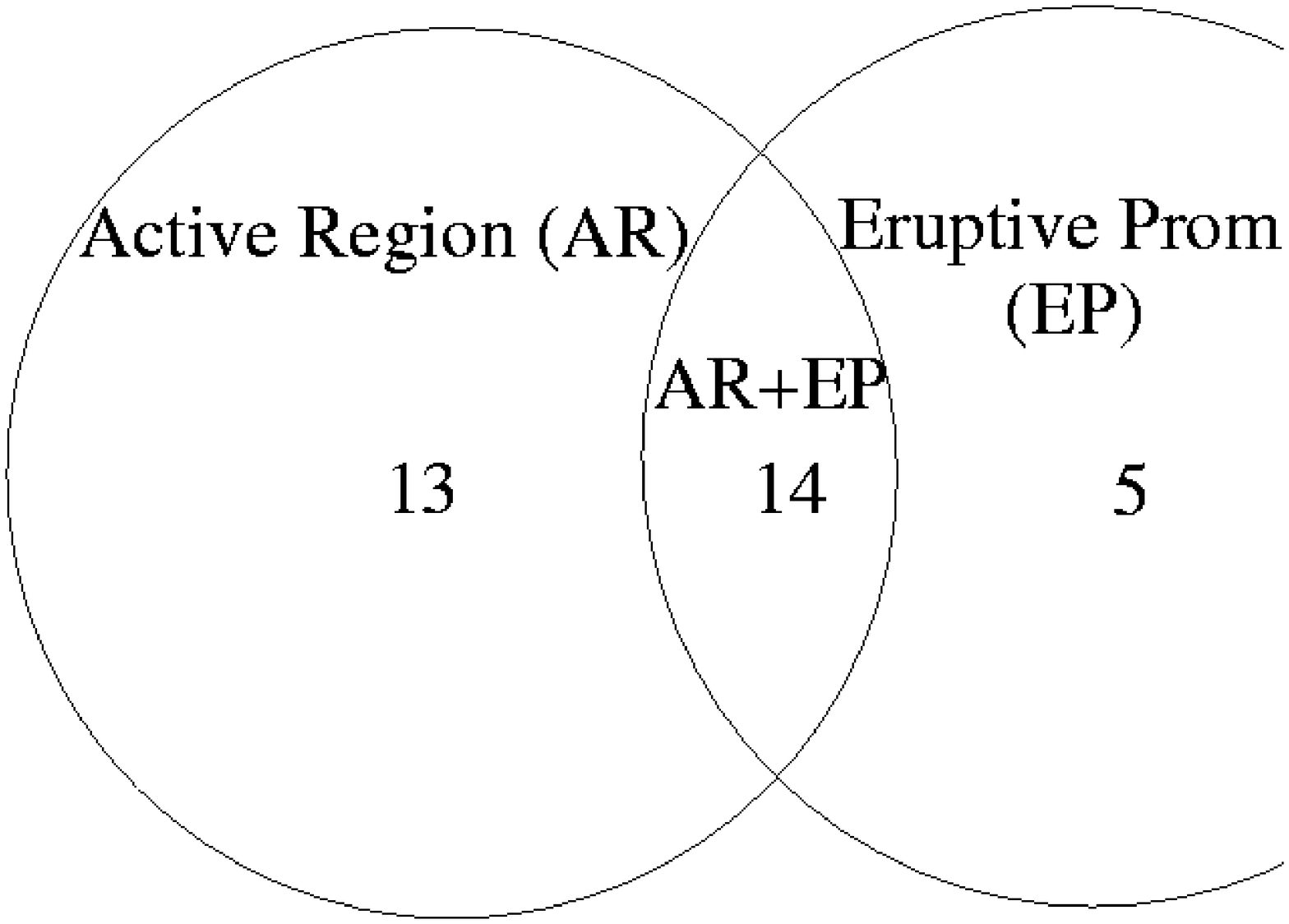]{This figure gives the number of on-disk CMEs in each of
the categories AR, AR+EP and EP. Of the 32 on-disk CMEs recorded in EIT 195 {\AA} 
data, 13 are associated with active regions (AR), 5 with eruptions
of prominences outside active regions (EP) and 14 with eruptions of active region
prominences (AR+EP).\label{f2}}

\figurenum{3a}
\figcaption[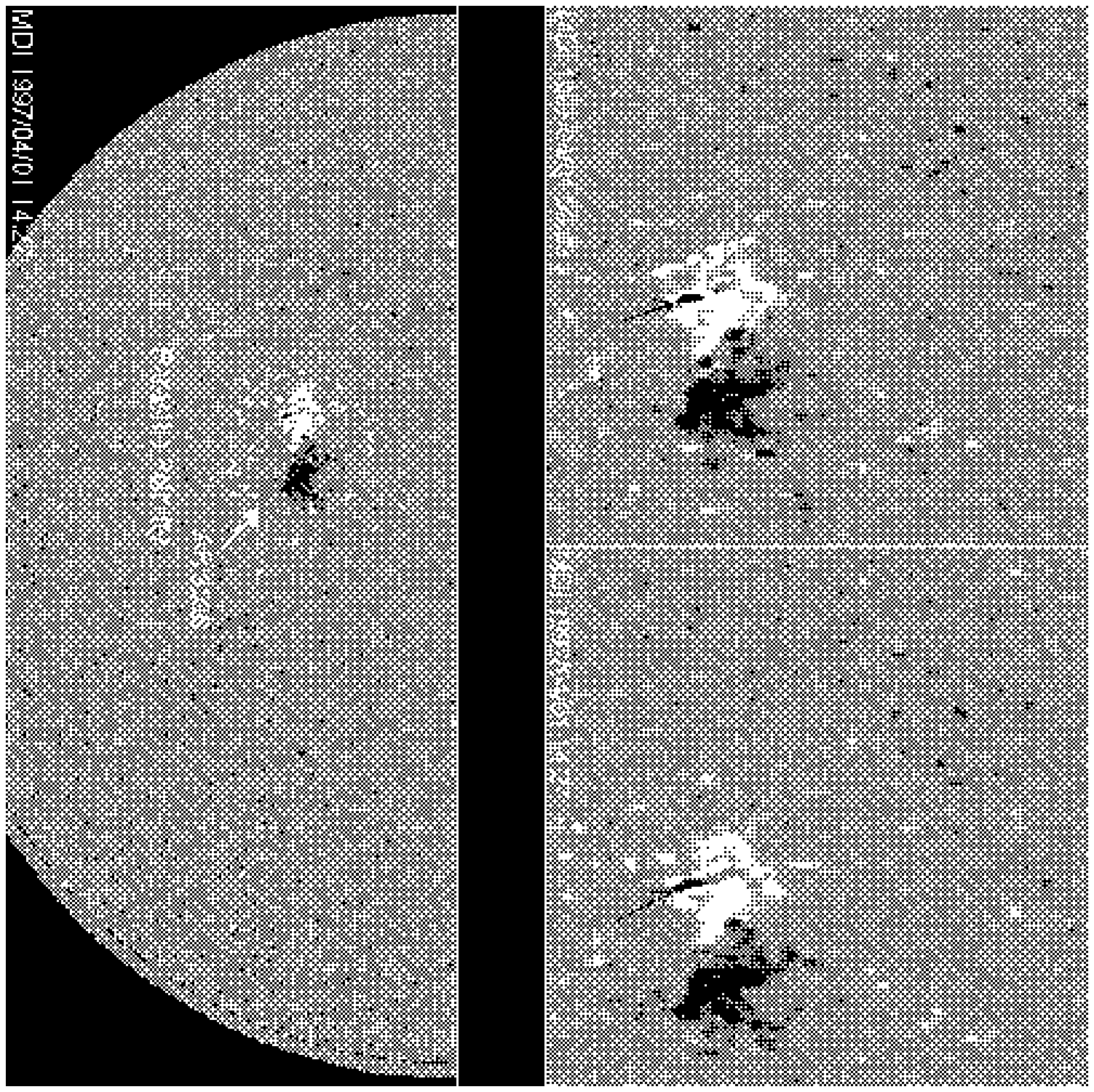]{MDI magnetogram data for 
AR 8026, which was the source region for the CME of 01 April 1997.
The bottom panel shows the location of the overall active region
on the solar disk, while the top panels show a closeup of the
active region. The parasitic negative polarity (marked by the
black arrow on the top panels) grows over a timescale of $\sim$ 
6 hours near the CME initiation time 
to form a lane in the midst of the positive polarity
region. The CME starts developing on the disk as seen in
EIT 195 {\AA} images around 13:46 UT, and the parasitic polarity
continues to grow and form the lane well beyond 14:27 UT.\label{f3a}}

\figurenum{3b}
\figcaption[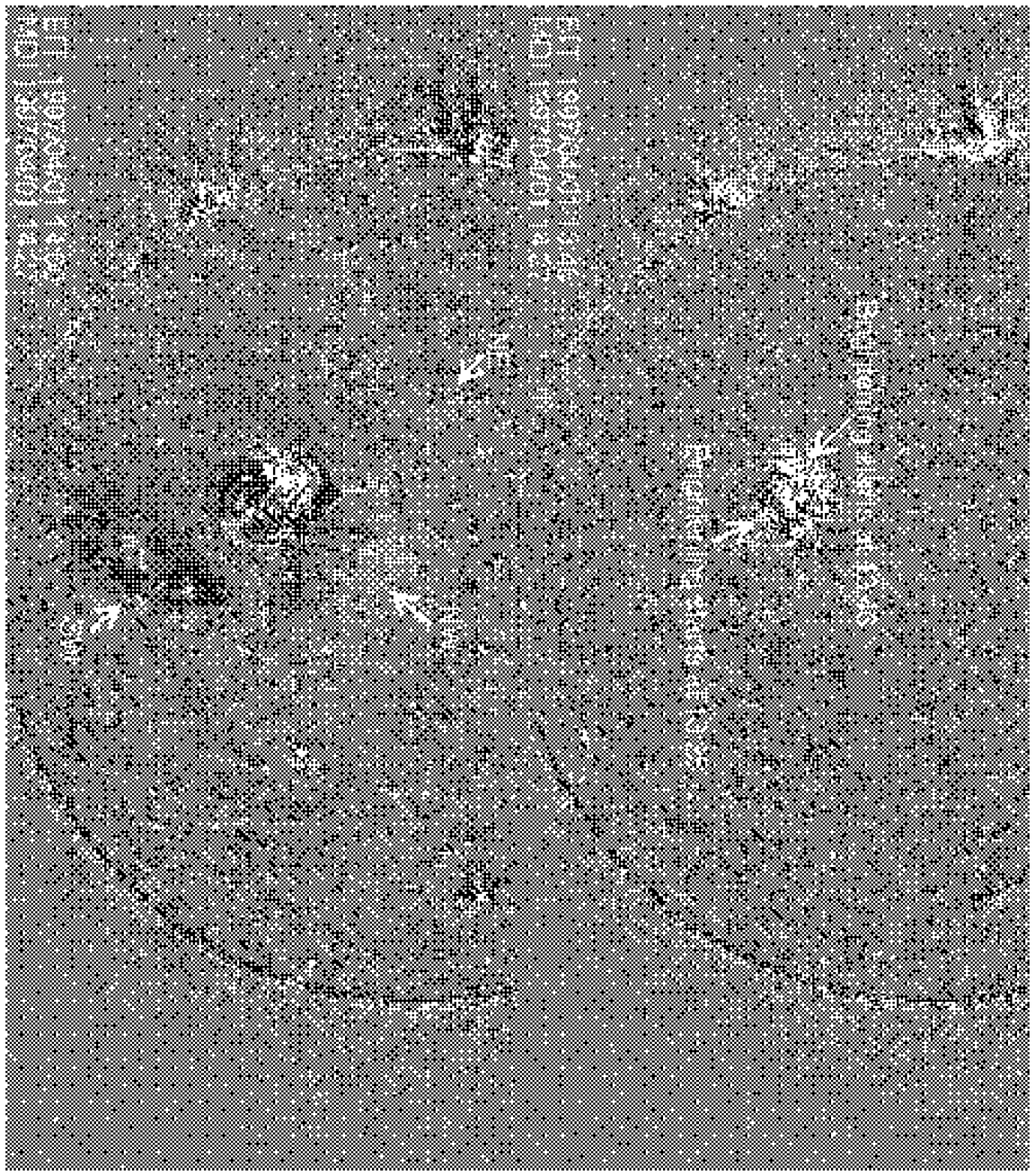]{Running difference
EIT 195 {\AA} images that show the development of the CME of 01 April
1997. Contours of the photospheric magnetic field from MDI data are
superimposed. The red contours denote +10 gauss fields, yellow +50
gauss, blue -10 gauss and green -50 gauss.
\label{f3b}}

\clearpage

\figurenum{3c}
\figcaption[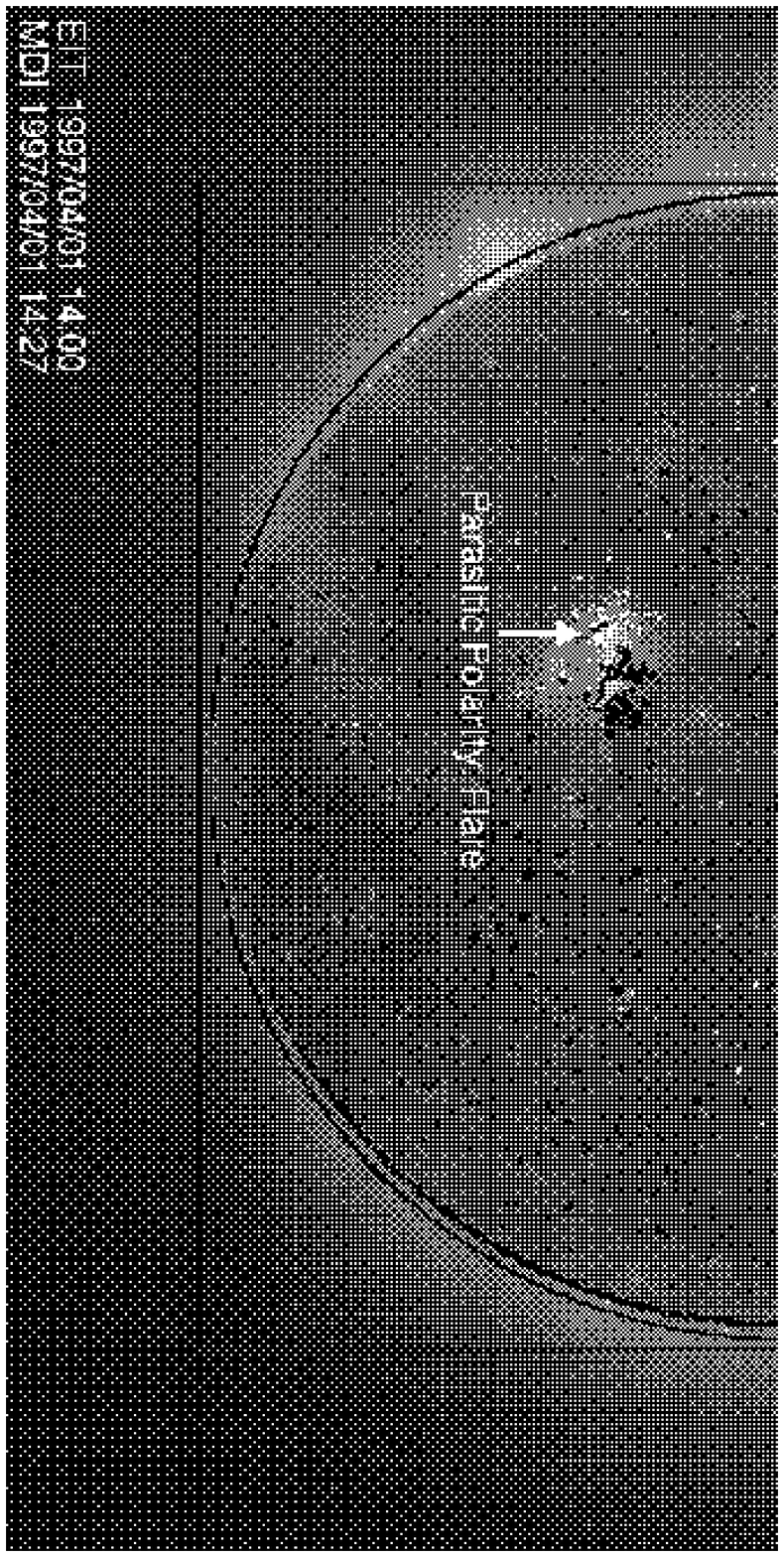]{Straight EIT 195 {\AA} image showing part
of the flare associated with the CME of 01 April 1997. 
Contours of the photospheric magnetic field from
MDI data are superimposed. The red contours denote +10 gauss fields,
blue -10 gauss, yellow +50 gauss and green -50 gauss.
It can be seen that the flare occurs at the location of the
parasitic negative polarity.\label{f3c}}

\figurenum{3d}
\figcaption[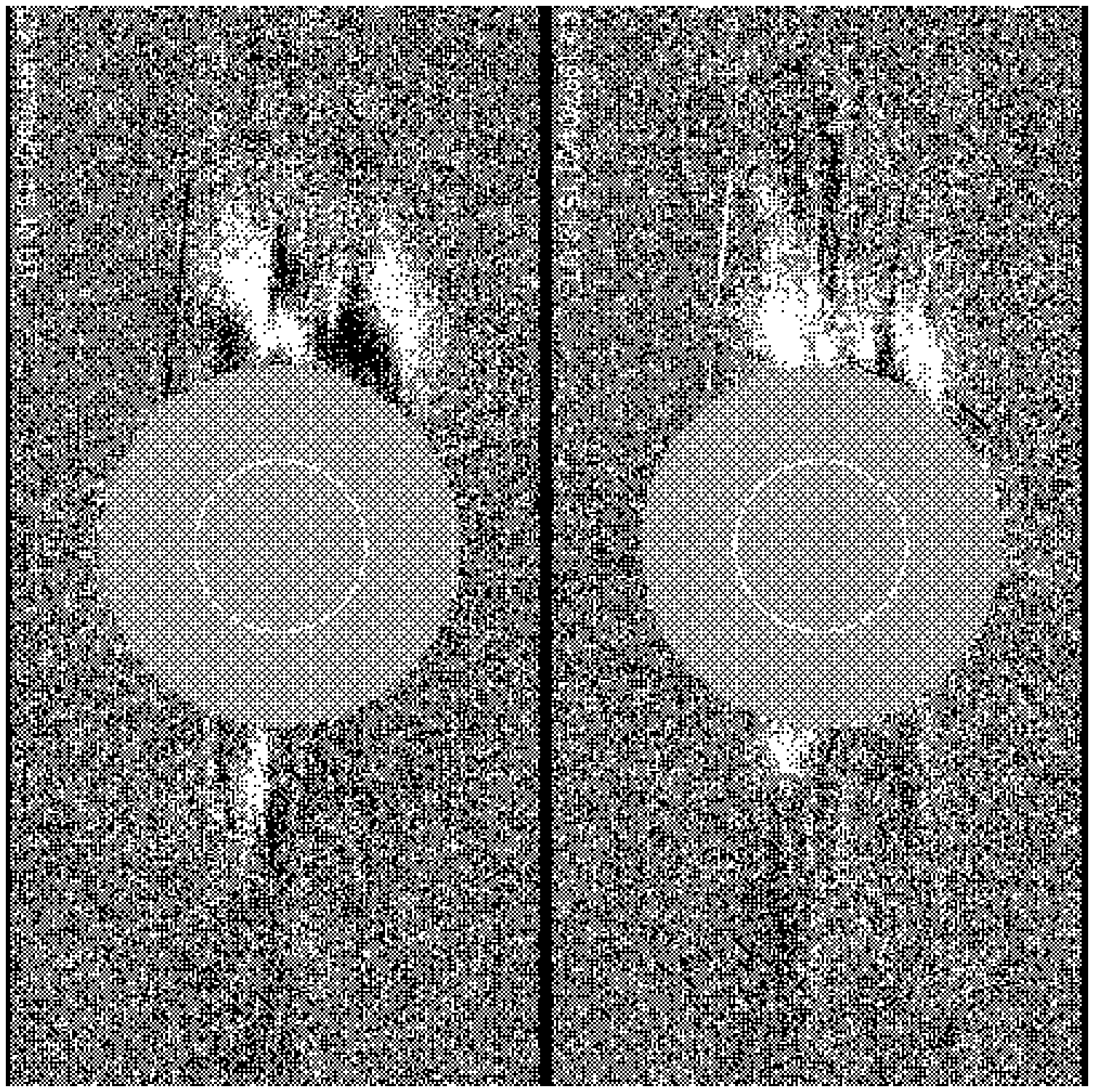]{Running difference images from the LASCO C2
coronograph showing the development of the CME of 01 April 1997.
The inner white circle in the center is at 1 $R_{\odot}$ while the
outer circle is the boundary of the C2 
occulter at 2.2 $R_{\odot}$.\label{f3d}}

\figurenum{4a}
\figcaption[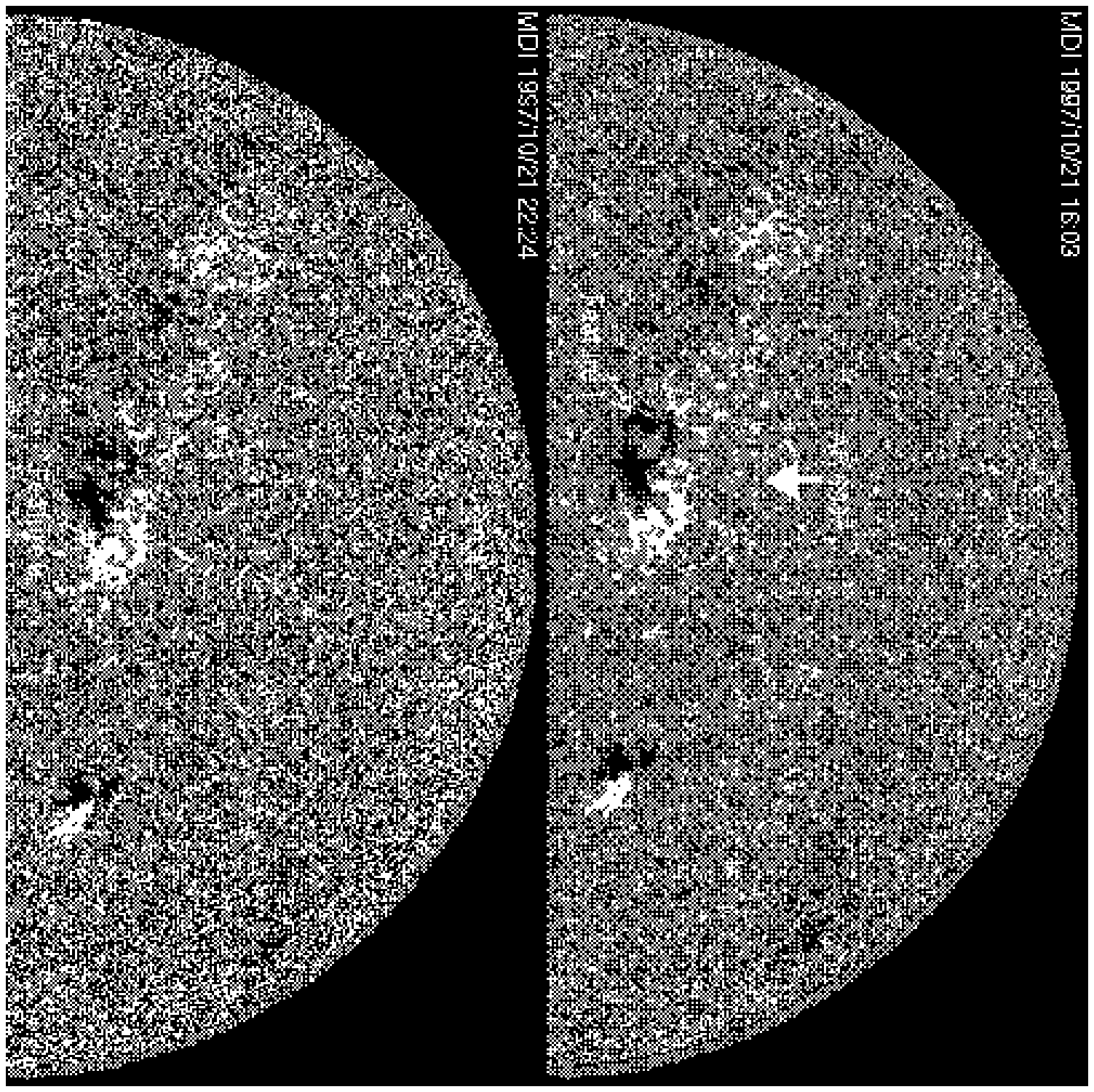]{MDI magnetogram data showing AR 8907 and
the associated decaying active region complex to its
northeast. The CME of 21 October 1997 was associated with
the eruption of a filament above the neutral line of the
decaying active region complex.\label{f4a}}

\figurenum{4b}
\figcaption[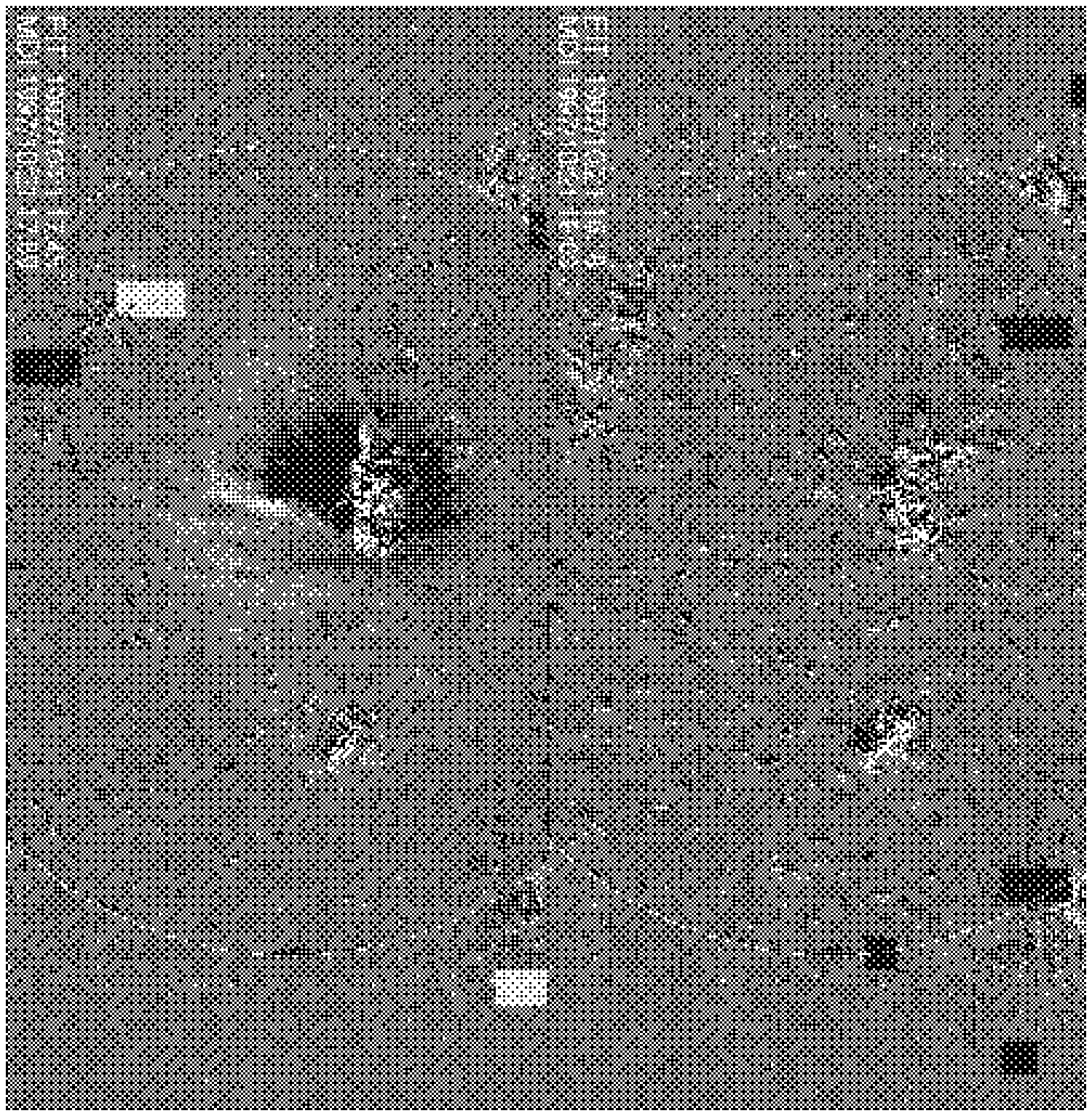]{Running difference
EIT 195 {\AA} images that show the development of the CME of
21 Oct 1997. Contours of the photospheric magnetic field from
MDI data are superimposed. The red contours denote +10 gauss fields,
blue -10 gauss, yellow +50 gauss and green -50 gauss.\label{f4b}}

\figurenum{4c}
\figcaption[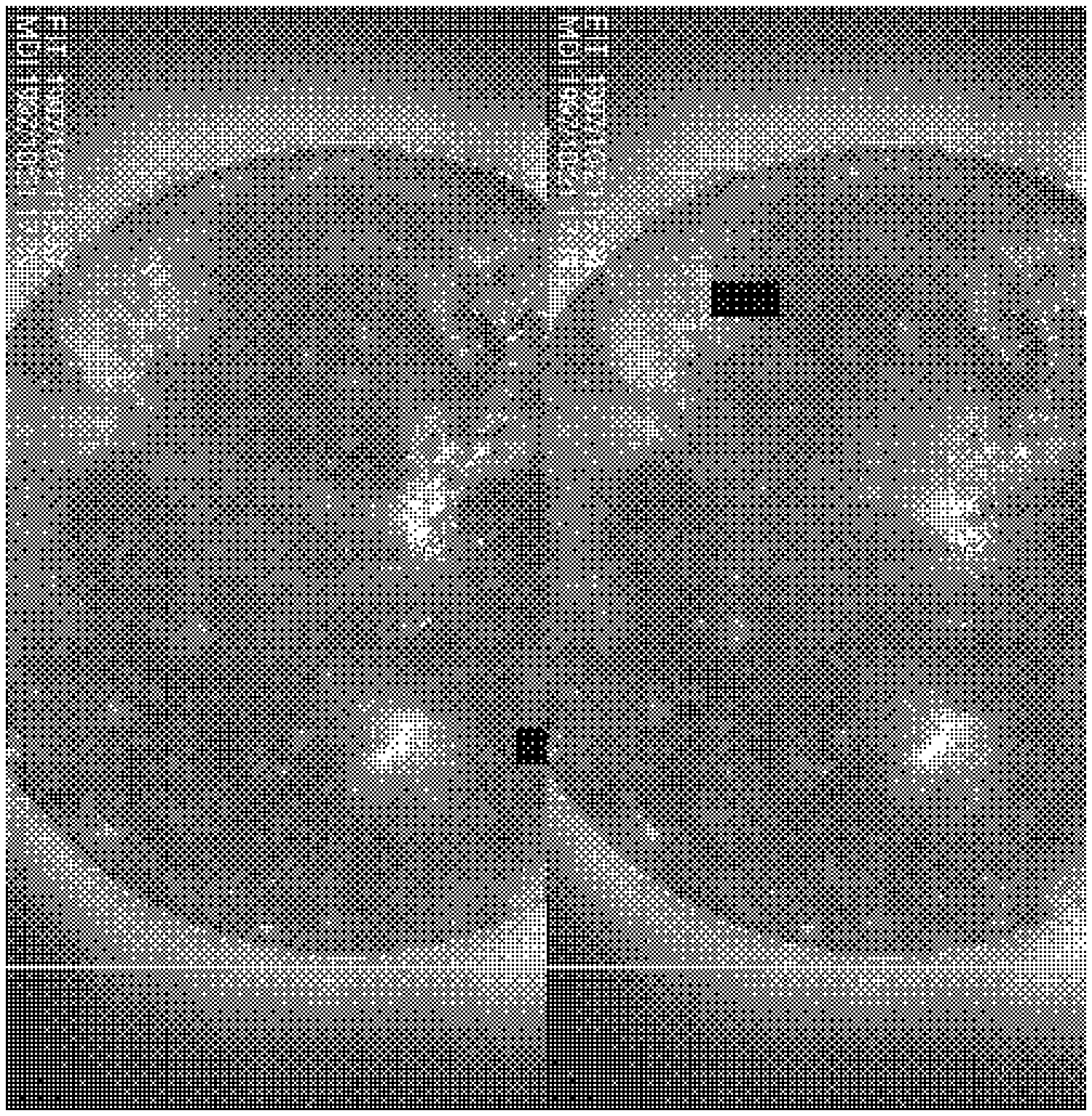]{Straight EIT 195 {\AA} images showing
the flaring activity asociated with the CME of 21 Oct 1997.
The twin ribbon configuration evident in the lower panel is
of special interest. 
Contours of the photospheric magnetic field from 
MDI data are superimposed. The
yellow contours denote +50 gauss fields and the
green contours -50 gauss fields.\label{f4c}}

\figurenum{4d}
\figcaption[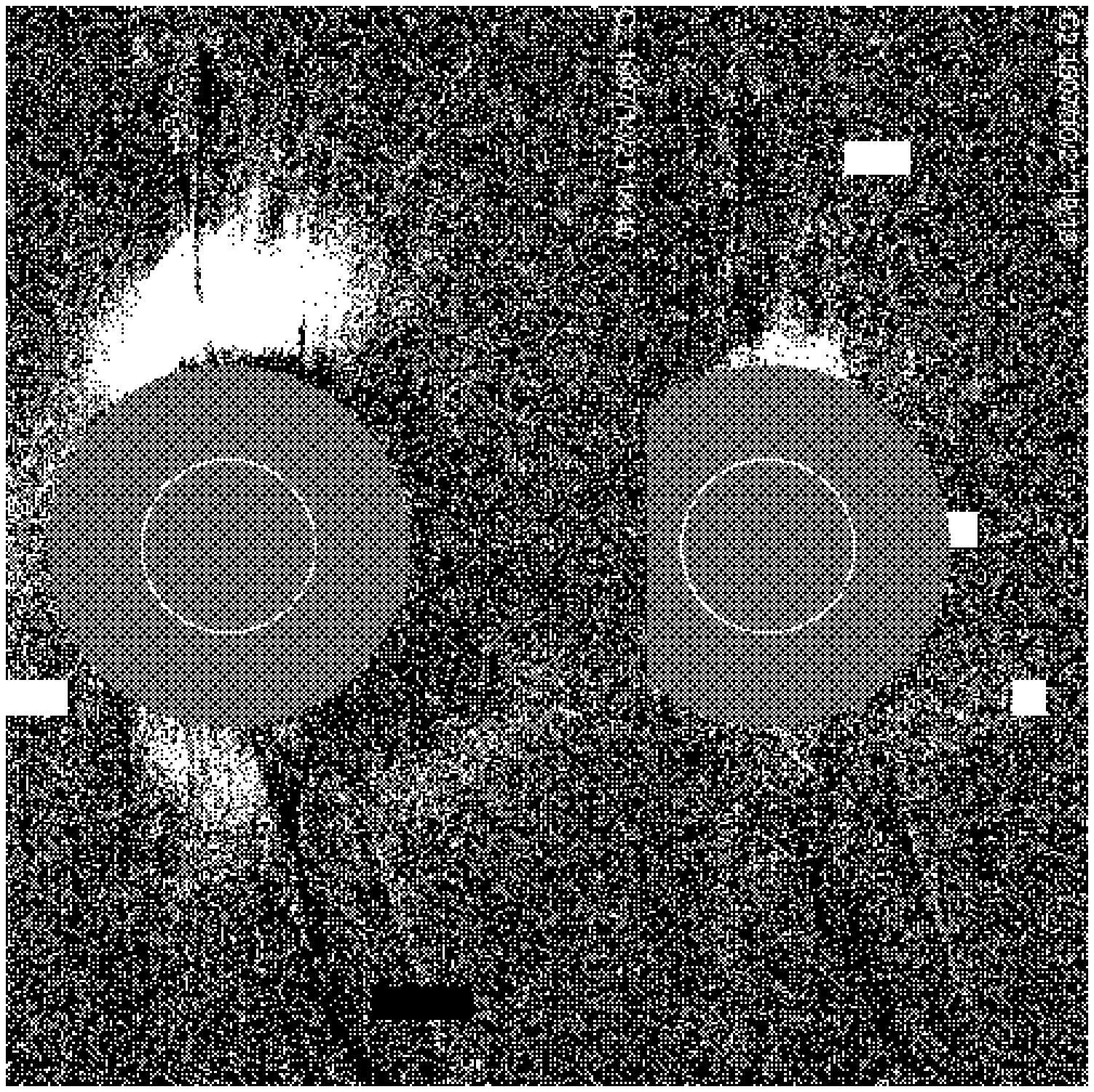]{Running difference images from the LASCO C2
coronograph showing the development of the CME of 21 Oct 1997.
The inner white circle in the center is at 1 $R_{\odot}$ while the
outer circle is the boundary of the C2 
occulter at 2.2 $R_{\odot}$.\label{f4d}}

\clearpage

\figurenum{5a}
\figcaption[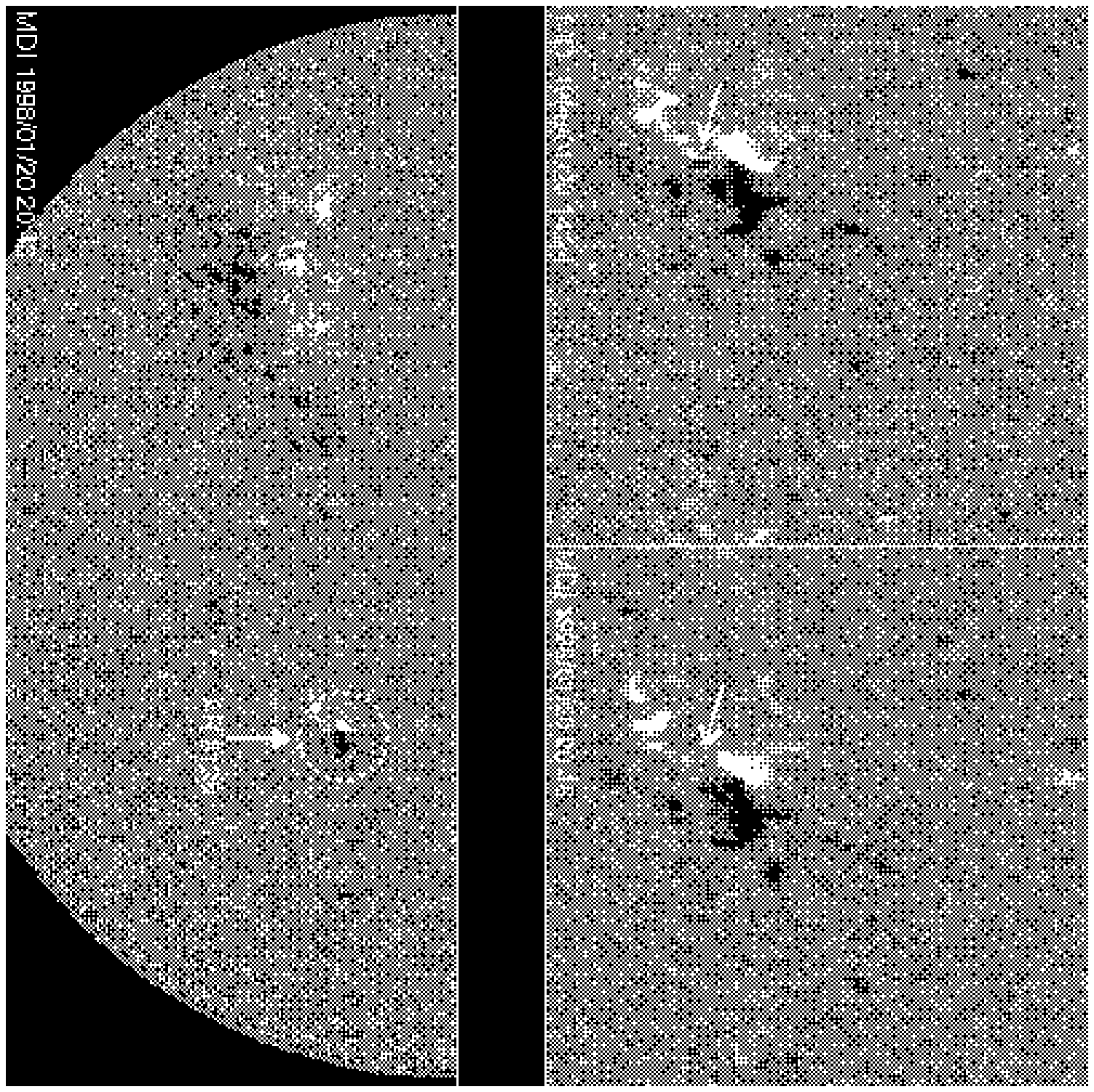]{MDI magnetogram data for 
AR 8135, which was the source region for the CME of 20 Jan 1998.
The bottom panel shows the location of the overall active region
on the solar disk, while the top panels show a closeup of the
active region. The top panels show flux cancellation near the area
marked by the white arrows over a $\sim$ 6 hour timescale 
before the initiation
of the CME in EIT 195 {\AA} data.
\label{f5a}}

\figurenum{5b}
\figcaption[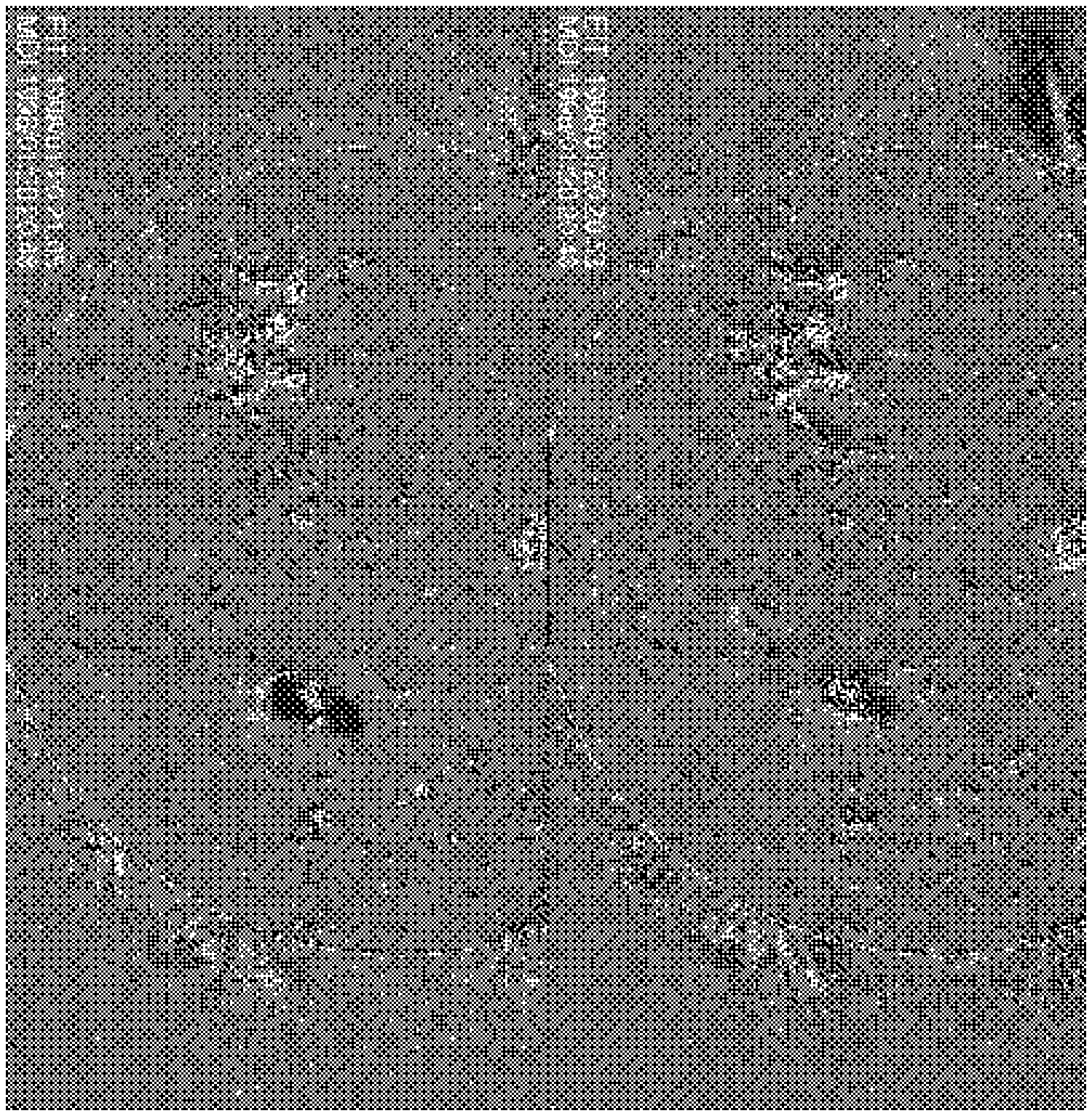]{Running difference
EIT 195 {\AA} images that show the development of the CME of
20 Jan 1998. Contours of the photospheric magnetic field from
MDI data are superimposed. The red contours denote +10 gauss fields,
blue -10 gauss, yellow +50 gauss and green -50 gauss.\label{f5b}}

\figurenum{5c}
\figcaption[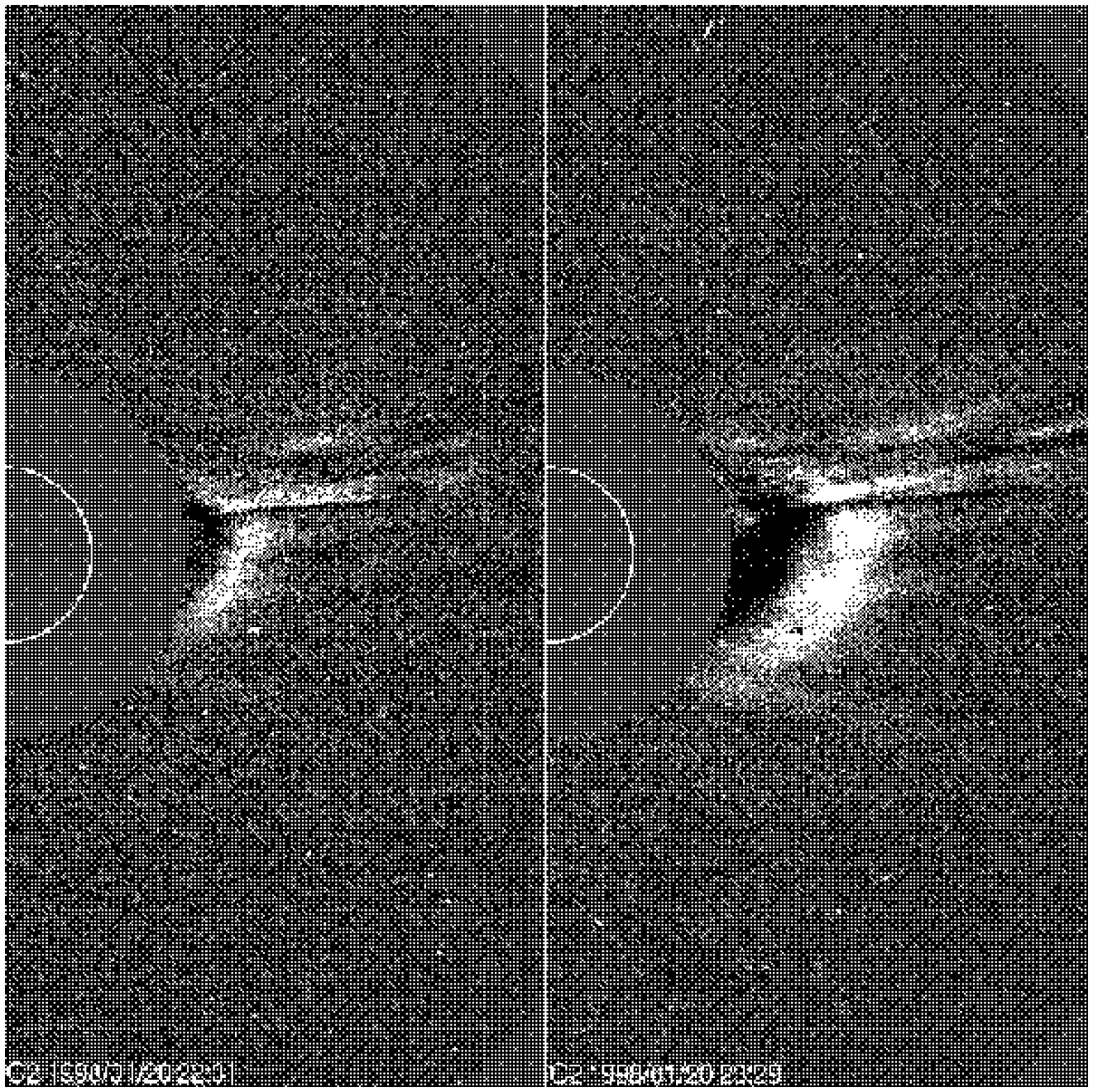]{Running difference images from the LASCO C2
coronograph showing the development of the CME of 20 Jan 1998.
The inner white circle in the center is at 1 $R_{\odot}$ while the
outer circle is the boundary of the C2 
occulter at 2.2 $R_{\odot}$.\label{f5c}}

\figurenum{6a}
\figcaption[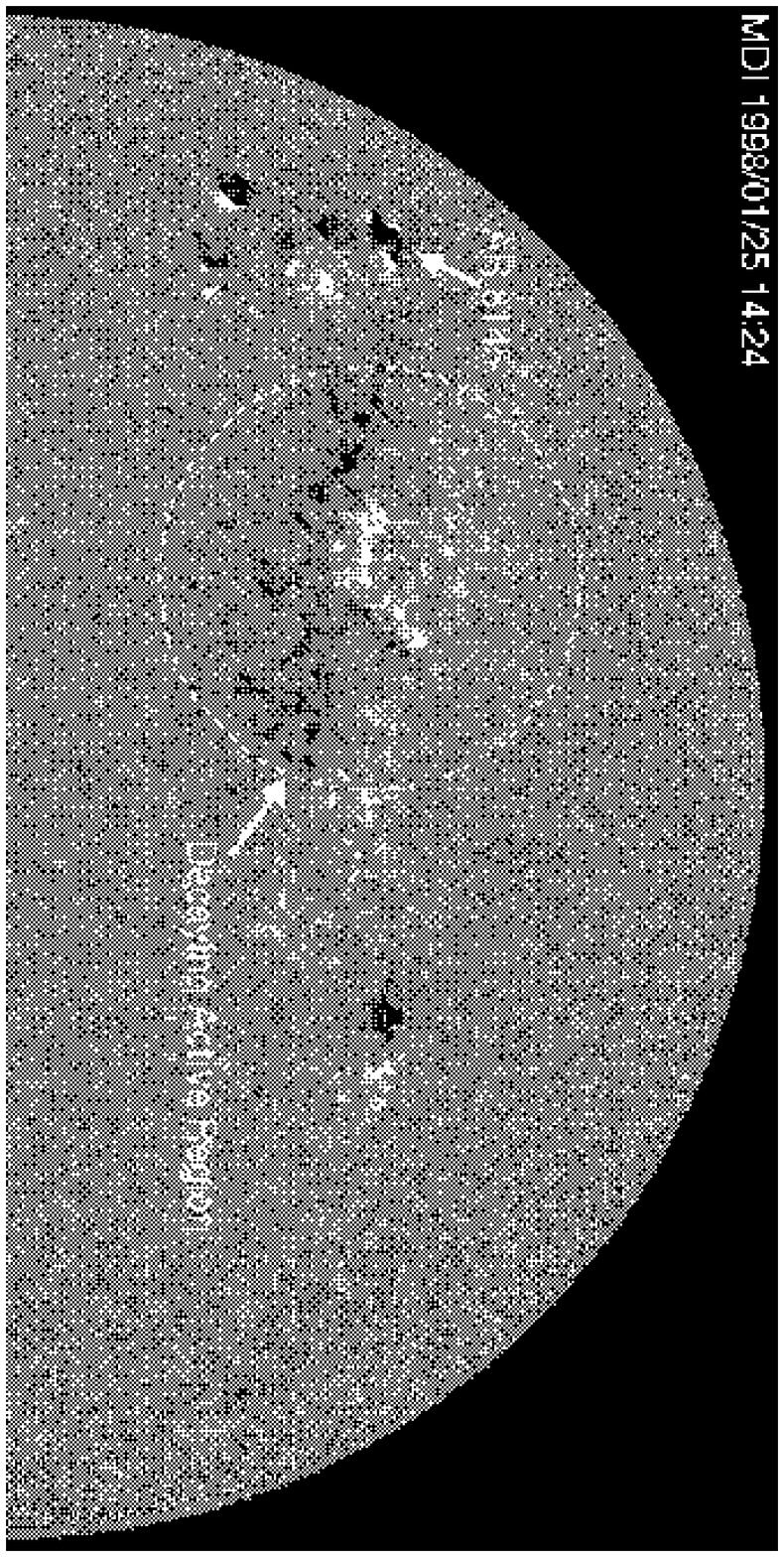]{MDI magnetogram data showing AR 8145 and
the associated decaying active region complex to its
west. The CME of 25 Jan 1998 was associated with
the eruption of a filament above the neutral line of the
decaying active region complex.\label{f6a}}

\figurenum{6b}
\figcaption[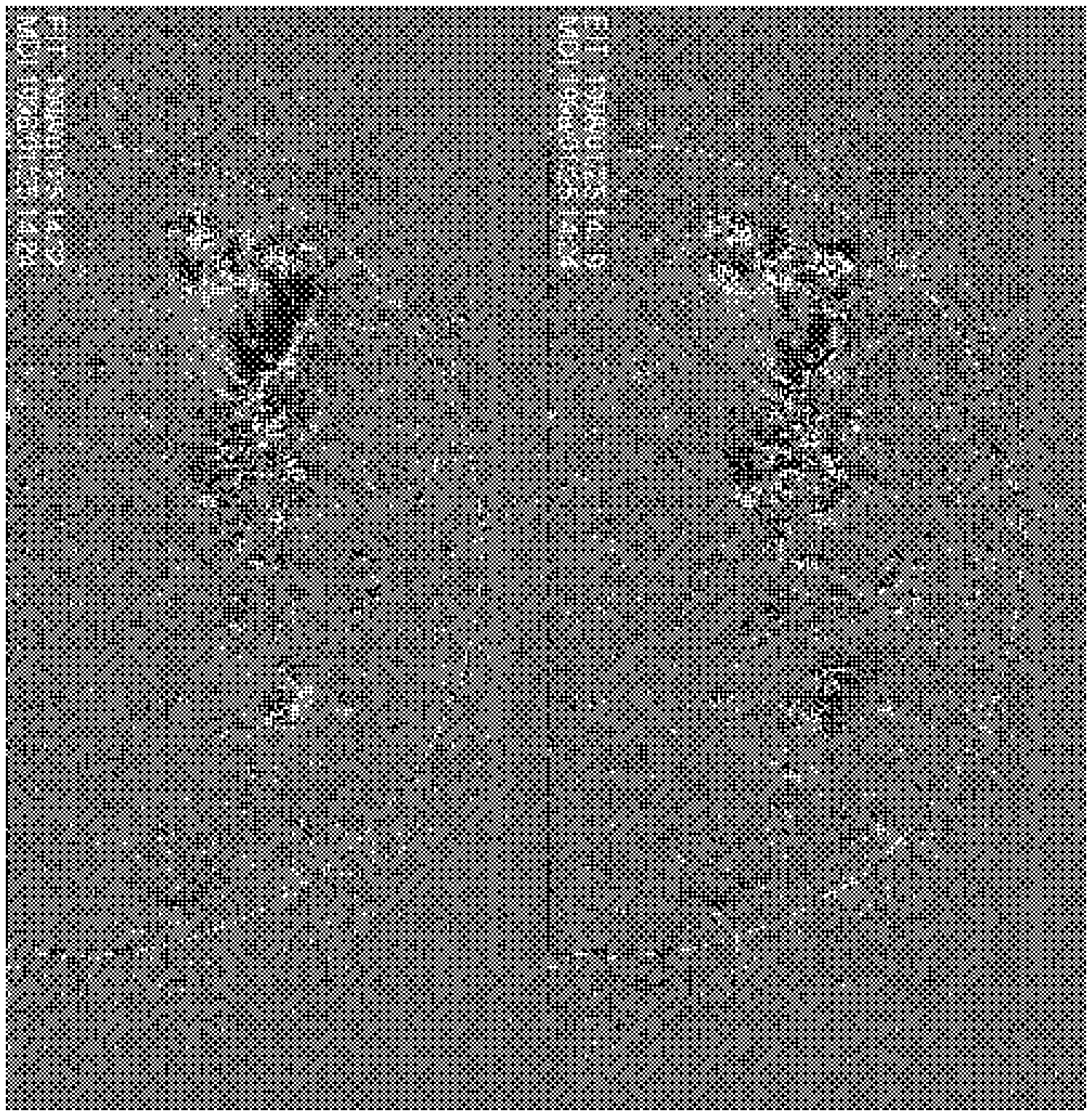]{Running difference
EIT 195 {\AA} images that show the development of the CME of
25 Jan 1998. Contours of the photospheric magnetic field from
MDI data are superimposed. The red contours denote +10 gauss fields,
blue -10 gauss, yellow +50 gauss and green -50 gauss.\label{f6b}}

\figurenum{6c}
\figcaption[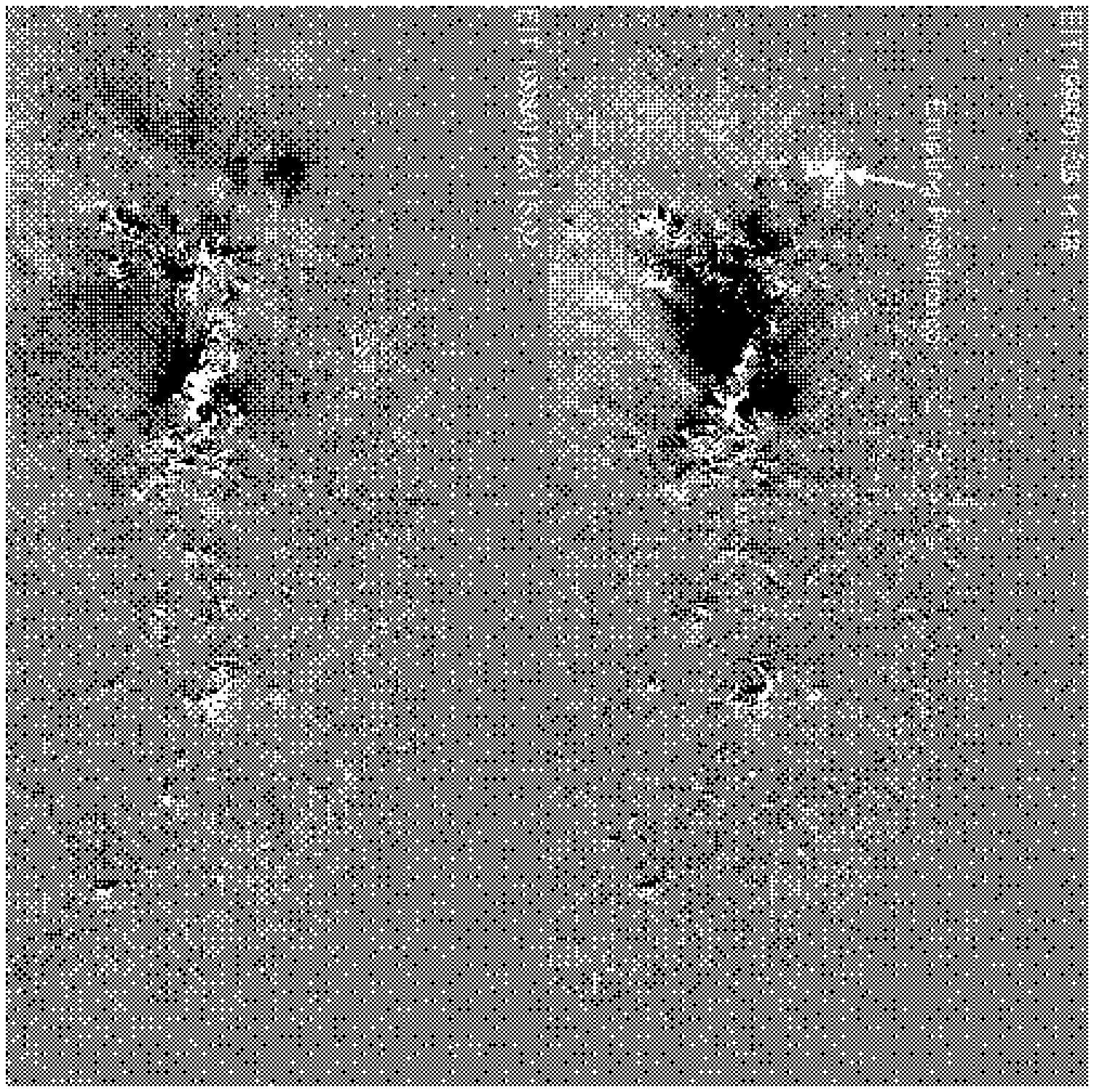]{Running difference EIT 195 {\AA} images
that show the latter stages of development of the CME of
25 Jan 1998.\label{f6c}}

\clearpage

\figurenum{6d}
\figcaption[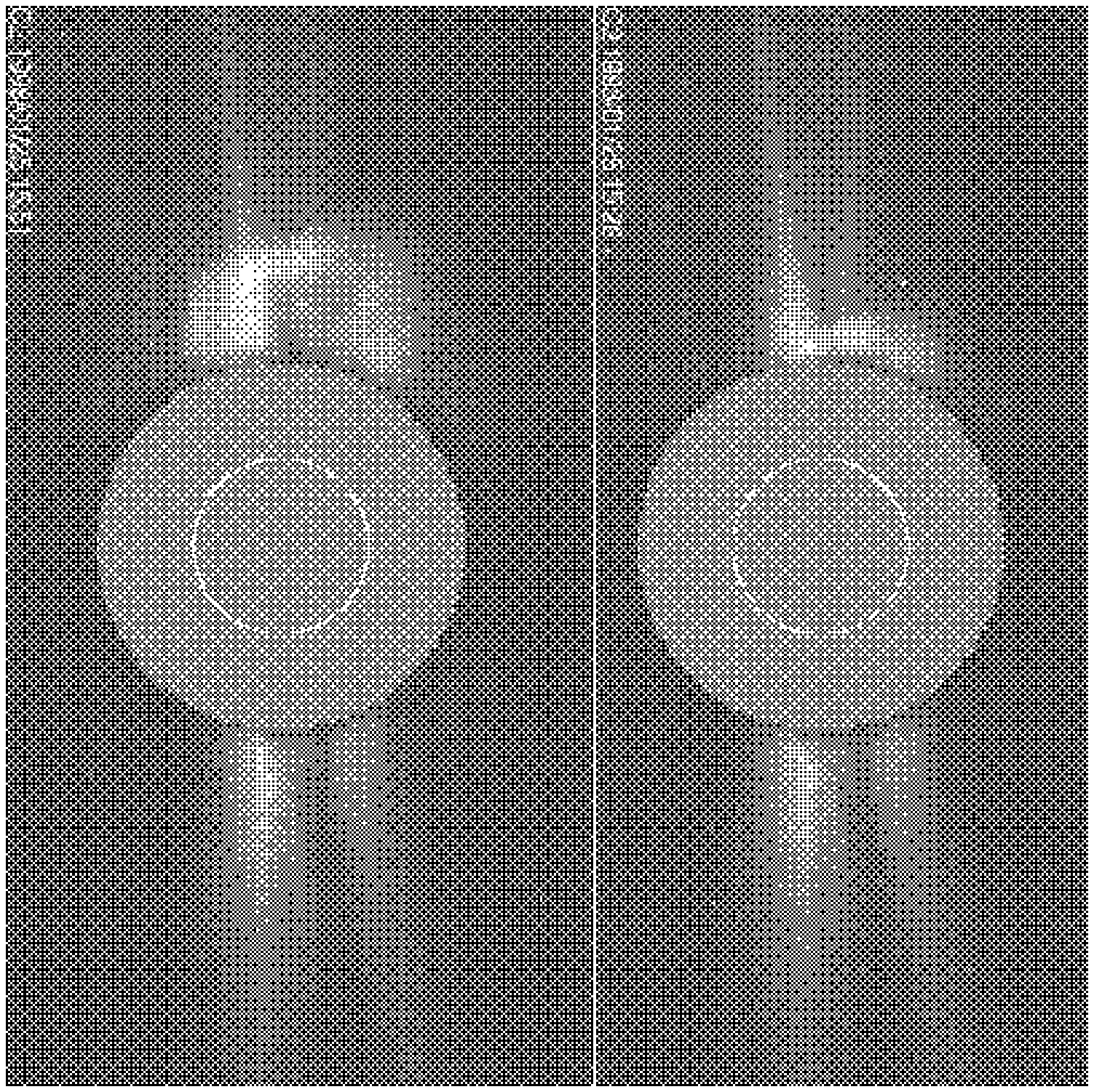]{Running difference images from the LASCO C2
coronograph showing the development of the CME of 25 Jan 1998.
The inner white circle in the center is at 1 $R_{\odot}$ while the
outer circle is the boundary of the C2 
occulter at 2.2 $R_{\odot}$.\label{f6d}}

\figurenum{7a}
\figcaption[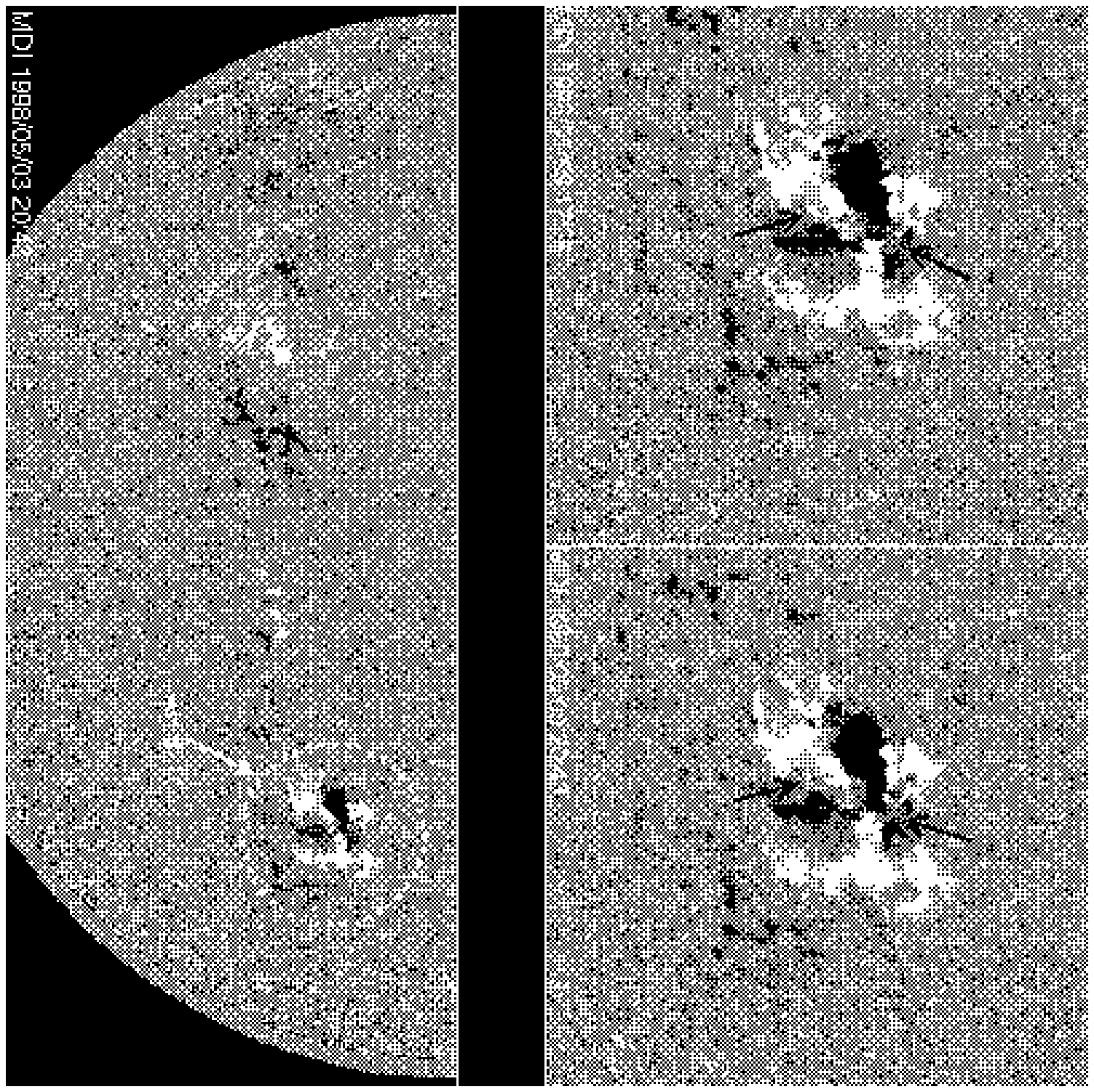]{MDI magnetogram data for 
AR 8210, which was the source region for the CME of 03 May 1998.
This active region is the source of several other CMEs and flares
(Table 3).
The bottom panel shows the location of the overall active region
on the solar disk, while the top panels show a closeup of the
active region. The top panel shows that 
there are two regions of cancelling magnetic
flux over a timescale of $\sim$ 6 hours before the initiation of
the CME in EIT 195 {\AA} data.
\label{f7a}}

\figurenum{7b}
\figcaption[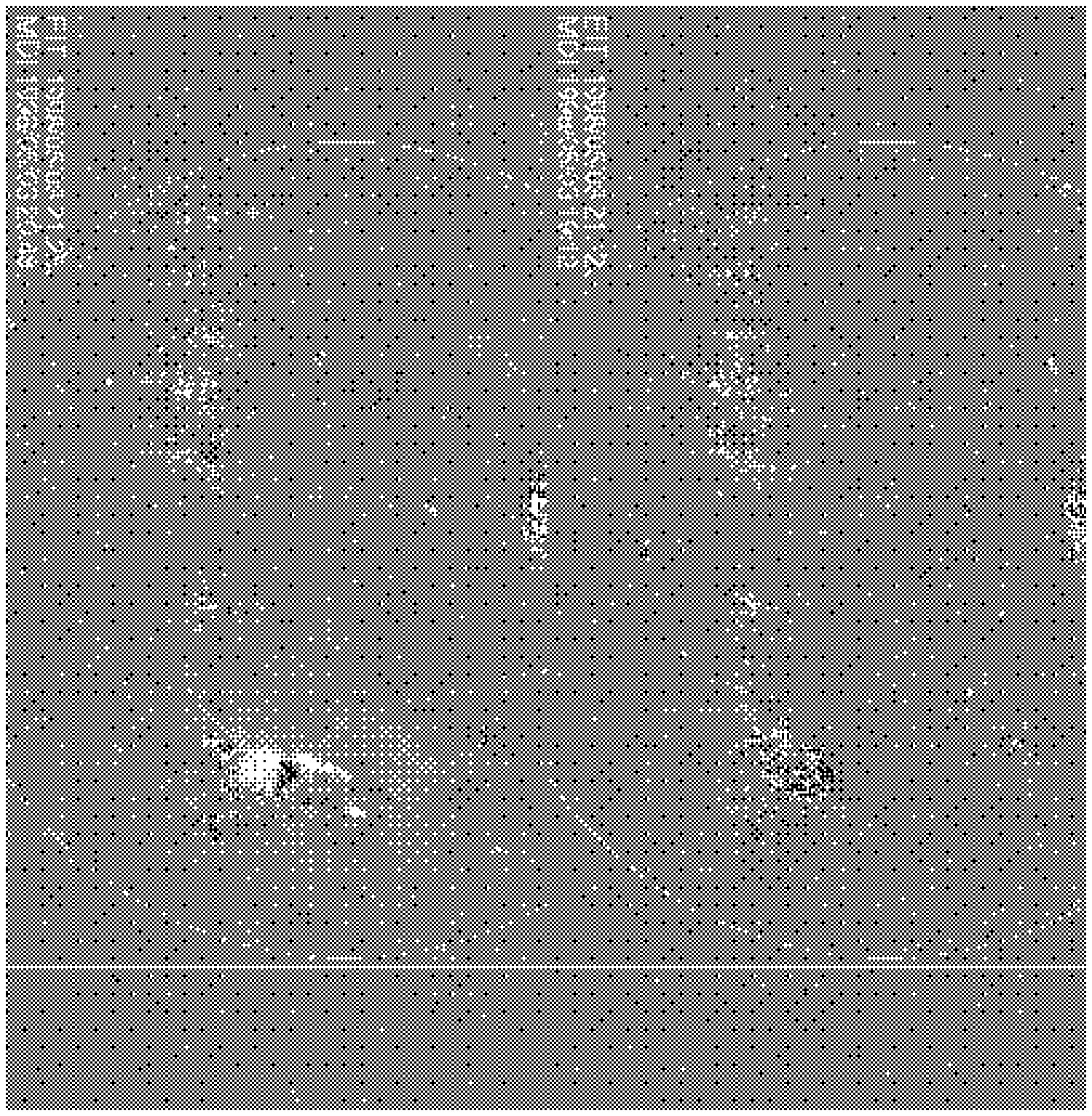]{Running difference
EIT 195 {\AA} images that show the development of the CME of
03 May 1998. Contours of the photospheric magnetic field from
MDI data are superimposed. The
yellow contours denote +50 gauss fields and the
green contours -50 gauss fields.\label{f7b}}

\figurenum{7c}
\figcaption[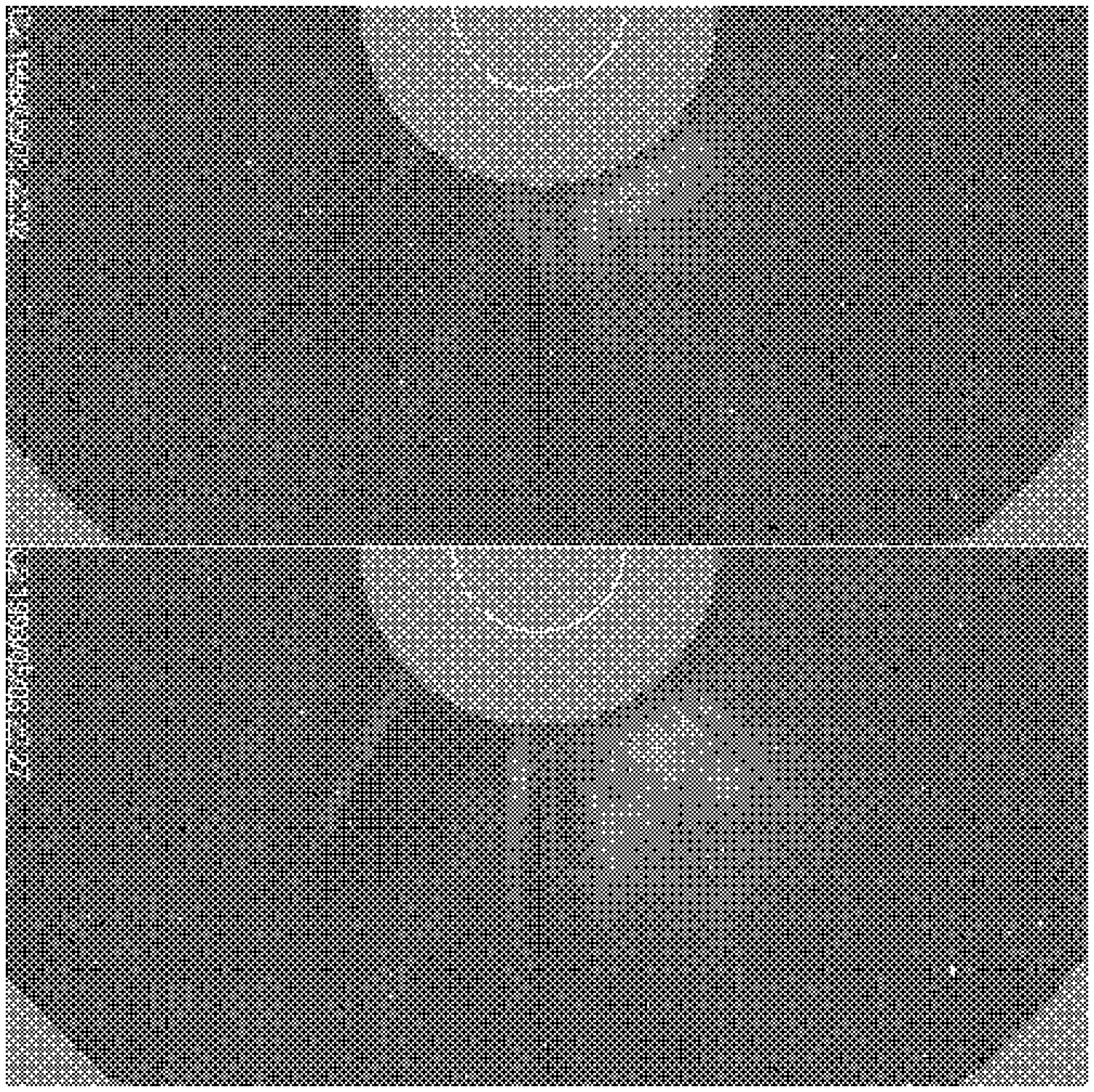]{Running difference images from the LASCO C2
coronograph showing the development of the CME of 03 May 1998.
The inner white circle in the center is at 1 $R_{\odot}$ while the
outer circle is the boundary of the C2 
occulter at 2.2 $R_{\odot}$.\label{f7c}}

\figurenum{8a}
\figcaption[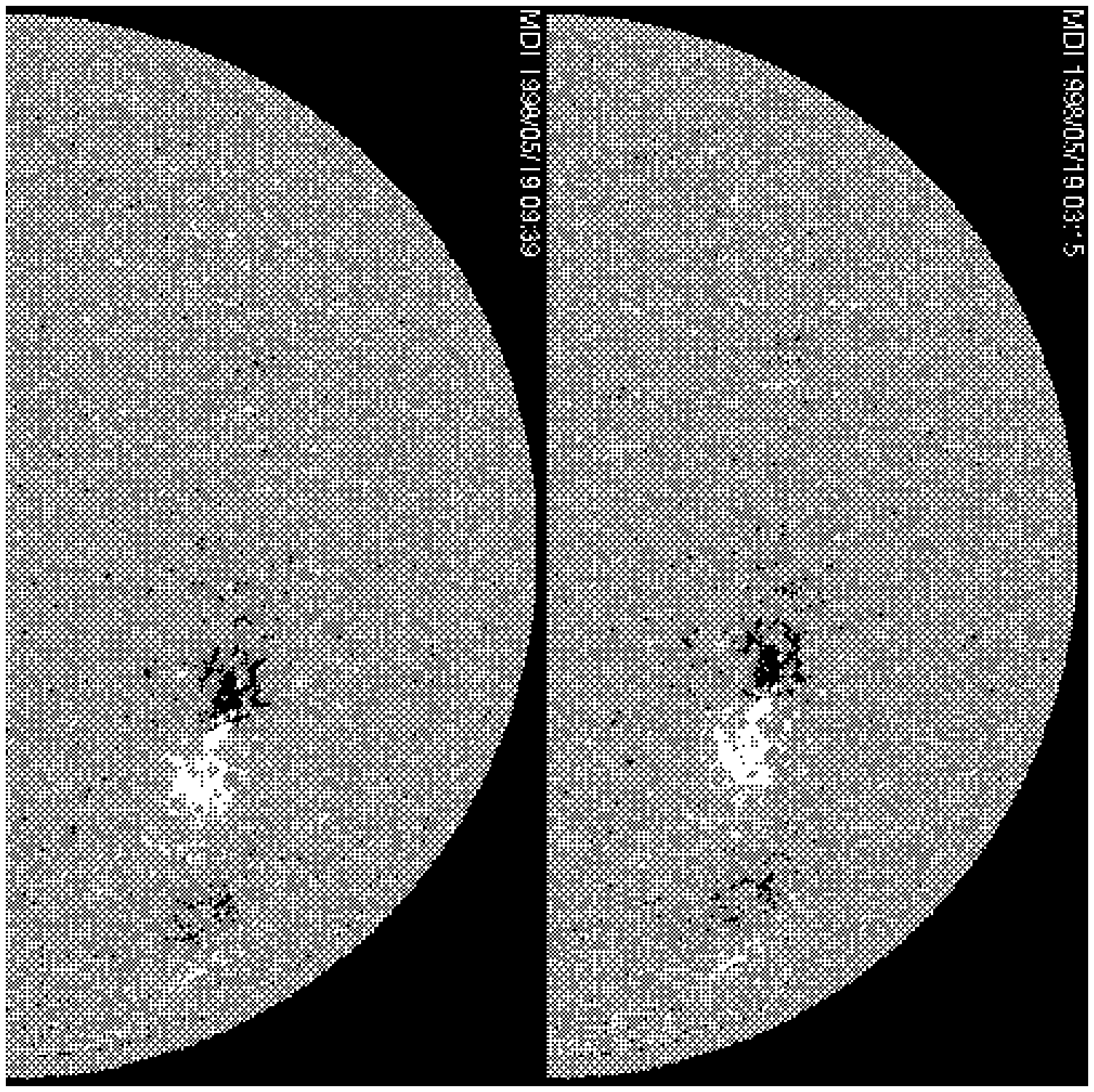]{MDI magnetogram data for 19 May 1998.
The prominence is embedded in the weak fields to the
west of the dominant active region.\label{f8a}}

\figurenum{8b}
\figcaption[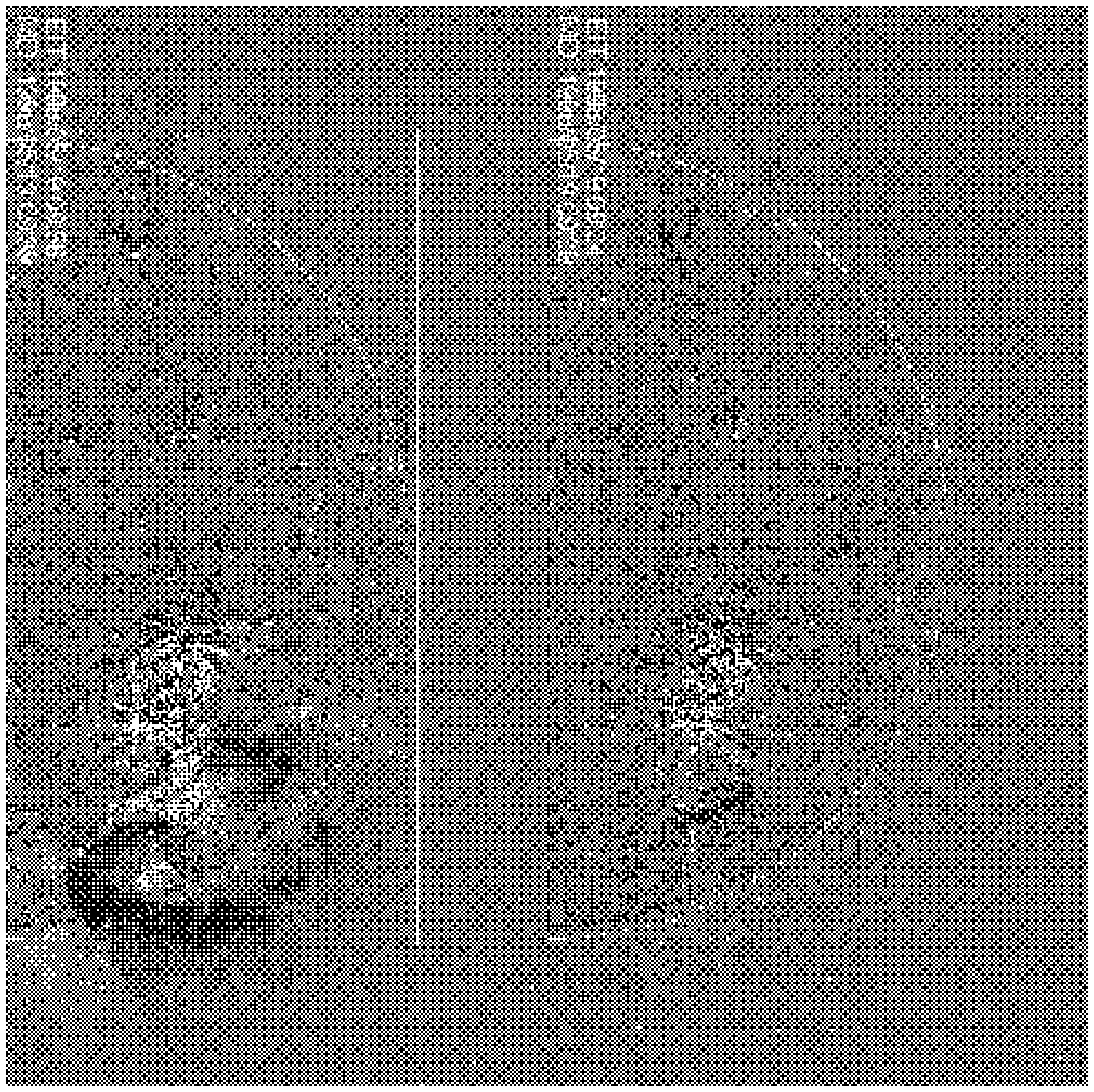]{Running difference
EIT 195 {\AA} images that show the prominence eruption of
19 May 1998. Contours of the photospheric magnetic field from
MDI data are superimposed. The red contours denote +10 gauss fields,
blue -10 gauss, yellow +50 gauss and green -50 gauss.\label{f8b}}

\clearpage

\figurenum{8c}
\figcaption[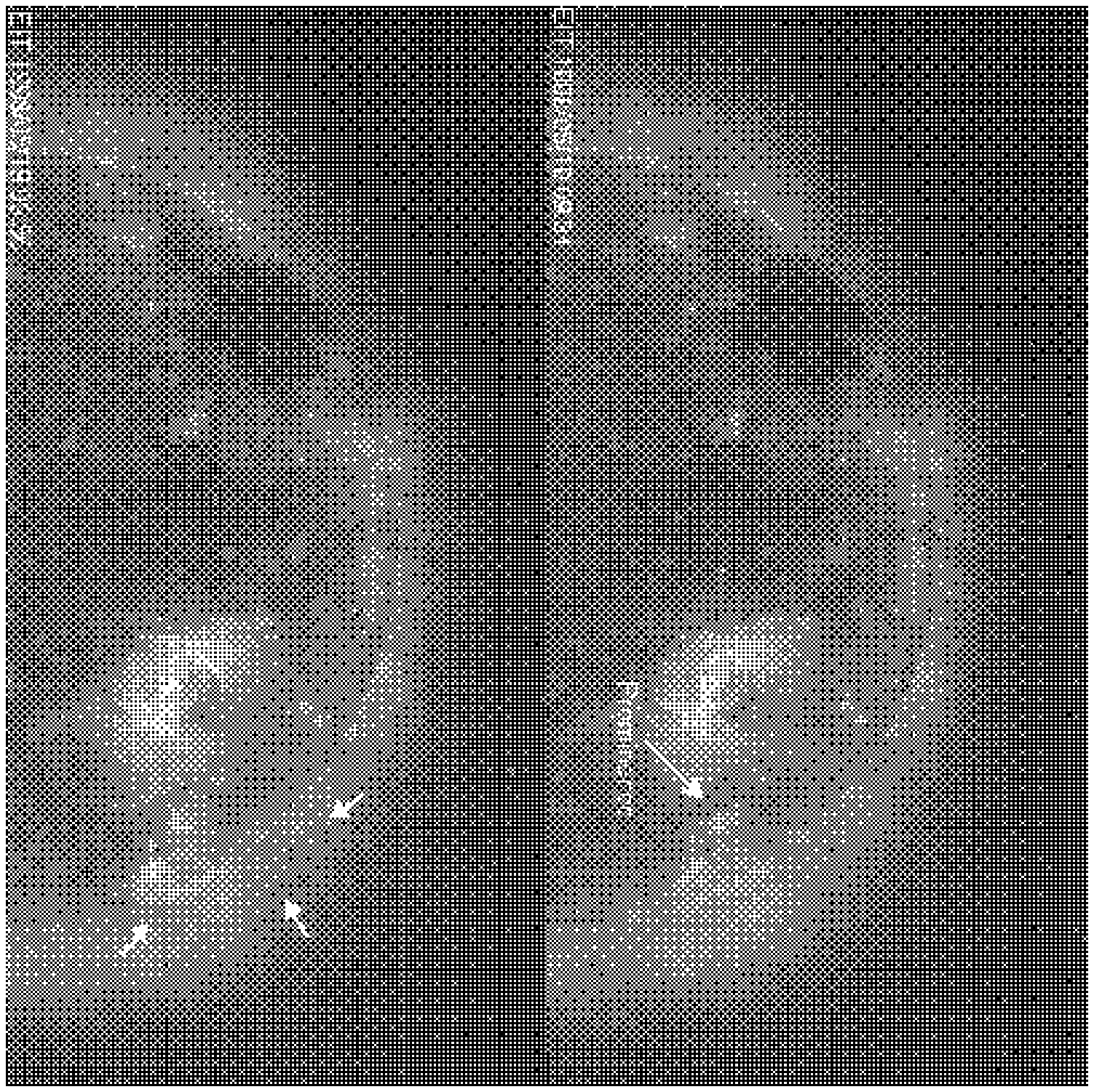]{Straight EIT 195 {\AA} images showing the 
prominence eruption of 19 May 1998.\label{f8c}}

\figurenum{8d}
\figcaption[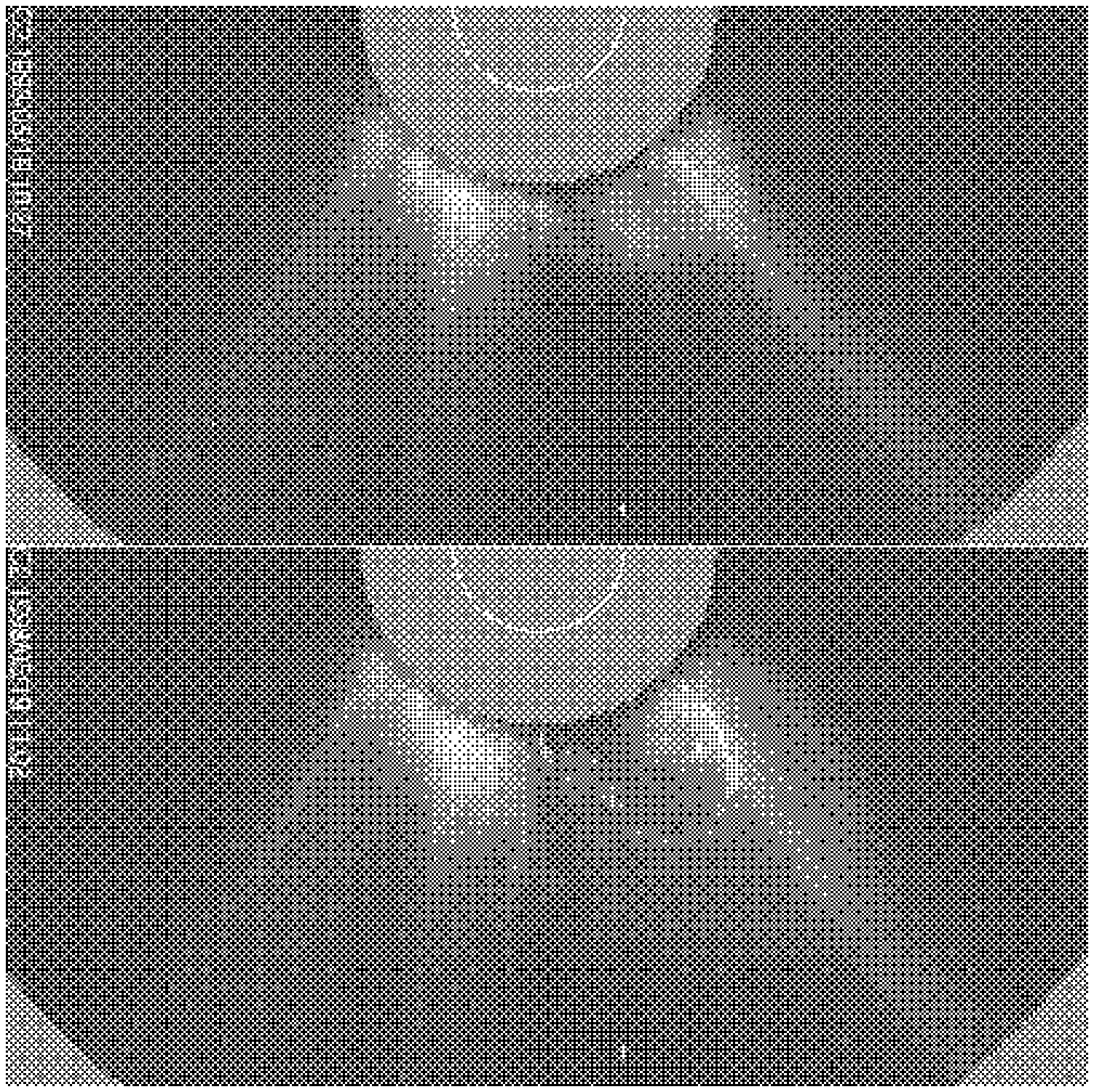]{White light images from the LASCO C2 coronograph
showing the development of the CME of 19 May 1998. The prominence
is clearly evident in the right panel. 
The inner white circle in the center is at 1 $R_{\odot}$ while the
outer circle is the boundary of the C2 
occulter at 2.2 $R_{\odot}$.\label{f8d}}





\clearpage

\begin{deluxetable}{ccc}
\footnotesize
\tablecaption{Catalog of all CMEs used in this study\label{tbl-3}}
\tablewidth{0pt}
\tablehead{
\colhead{Date} & \colhead{Initiation time (UT)\tablenotemark{1}}& Category} 
\startdata
97/04/01 & 13:46 & AR\tablenotemark{3}\\
97/10/10 & 03:45 & AR\\
97/10/11 & 08:51 & AR\\
97/11/04 & 05:58 & AR\\
97/11/06 & 11:41 & AR\\
98/01/08 & 08:59 & AR\\
98/01/17 & 23:36 & AR\\
98/01/20 & 21:05 & AR\tablenotemark{3}\\
98/01/27 & 22:20 & AR\\
98/04/27 & 08:51 & AR\tablenotemark{2}\\
98/05/06 & 07:59 & AR\tablenotemark{2}\\
98/05/27 & 11:19 & AR\\
98/06/08 & 16:04 & AR\\
96/12/23 & 20:20 & AR+EP\\
97/04/07 & 14:00 & AR+EP\\
97/05/12 & 04:50 & AR+(EP)\tablenotemark{4}\\
97/10/13 & 08:34 & AR+EP\\
97/10/21 & 17:34 & AR+EP\\
97/11/27 & 13:12 & AR+(EP)\tablenotemark{4}\\
98/01/25 & 14:32 & AR+EP\tablenotemark{3}\\
98/01/26 & 23:05 & AR+(EP)\tablenotemark{4}\\
98/03/27 & 00:45 & AR+(EP)\tablenotemark{4}\\
98/04/15 & 07:14 & AR+EP\\
98/04/29 & 15:52 & AR+EP\\
98/05/02 & 13:21 & AR+(EP)\tablenotemark{2,4}\\
98/05/03 & 21:12 & AR+EP\tablenotemark{2,3}\\
98/05/05 & 23:36 & AR+EP\tablenotemark{2}\\
96/10/19 & 15:00 & EP\\
97/10/06 & 14:10 & EP\\
97/10/21 & 00:20 & EP\tablenotemark{3}\\
98/01/21 & 06:43 & EP\\
98/05/19 & 07:54 & EP\tablenotemark{3}\\
\enddata

\tablenotetext{1}{Initiation time as discerned from the source region in
EIT 195 {\AA} data}
\tablenotetext{2}{Events associated with NOAA active region 8210}
\tablenotetext{3}{Case studies examined in detail (\S 3.2)}
\tablenotetext{4}{Evidence for an erupting prominence is not very strong}

\end{deluxetable}

\clearpage

\begin{deluxetable}{cccccc}
\footnotesize
\tablecaption{CMEs used for detailed case studies\label{tbl-2}}
\tablewidth{0pt}
\tablehead{
\colhead{CME} & \colhead{Category}   & Flare &
\colhead{AR lifetime} & \colhead{AR phase} & 
\colhead{Short Timescale Behavior ($\sim$ 6 hrs)} 
}
\startdata
97/04/01 & AR & M1.9 & 53-75 days & Rising phase & Parasitic polarity forms lane
\\
97/10/21 & AR+P & C3.3 & $\sim$ 6 months & Decaying & None \\
98/01/20 & AR & None & 11-36 days & Mid-life & Cancelling flux\\
98/01/25 & AR+P & C1.1 & $\sim$ 7 months & Decaying & None\\
98/05/03\tablenotemark{1} & AR & M1.4 & 
$\sim$ 65-79 days & Mid-life & Cancelling flux\\

\enddata

\tablenotetext{1}{Prolific active region, see Table 3}

\end{deluxetable}

\clearpage

\begin{deluxetable}{cccc}
\footnotesize
\tablecaption{Events associated with NOAA active region 8210\label{tbl-3}}
\tablewidth{0pt}
\tablehead{
\colhead{Date} & \colhead{CME in C2 (UT)}   & 
\colhead{Flare peak time (UT)} & \colhead{Flare class} 
}
\startdata
98/04/27 & 09:26 & 09:20 & X1.0 \\
98/05/02 & 14:06 & 13:42 & X1.1 \\
98/05/03 & 21:31 & 21:29 & M1.4 \\
98/05/06 & 00:02 & 23:46 (05/05) & M2.5 \\
98/05/06 & 08:04 & 08:09 & X2.7
\enddata
\end{deluxetable}

\clearpage

\newpage

\tableofcontents

\newpage

\listoffigures

\end{document}